\pdfoutput=1
\documentclass{article}
\usepackage{amsfonts}
\usepackage{amsmath}

\usepackage{pdfpages}

\input{tcilatex}
\begin{document}

\title{Characterizing Planetary Orbits and the Trajectories of Light}
\author{F.T. Hioe* and David Kuebel \and Department of Physics, St. John
Fisher College, Rochester, NY 14618 \and and \and Department of Physics \&
Astronomy, University of Rochester, Rochester, NY 14627}
\maketitle

\begin{abstract}
Exact analytic expressions for planetary orbits and light trajectories in
the Schwarzschild geometry are presented. A new parameter space is used to
characterize all possible planetary orbits. Different regions in this
parameter space can be associated with different characteristics of the
orbits. The boundaries for these regions are clearly defined. Observational
data can be directly associated with points in the regions. A possible
extension of these considerations with an additional parameter for the case
of Kerr geometry is briefly discussed.

\end{abstract}

\section{Introduction}

Nearly a century after Einstein's theory of general relativity was found to
correctly predict the precession of the planet Mercury around the sun and
the deflection of light by the sun's gravitational field, the problem of
understanding orbital trajectories around very massive objects still retains
interest as it relates to current astrophysical topics [1] such as the study
of gravitational waves. Among the numerous works on this subject, we should
mention the classic publications of Whittaker [2], Hagihara [3] and
Chandrasekhar [4] on the Schwarzschild geometry, and the more recent work of
Levin and Perez-Giz [5] on the Schwarzschild and Kerr geometries.

In the work of Chandrasekhar and that of Hagihara, the orbits are classified
into various types according to the roots of a certain cubic equation, while
in the work of Levin and Perez-Giz, the orbits are classified topologically
by a triplet of numbers that indicate the numbers of zooms, whirls and
vertices. In the work of Levin and Perez-Giz, the orbits were obtained by
numerically integrating the integrable equations. These authors used the
planet's energy and angular momentum as the principal physical parameters,
and made extensive use of an effective potential for describing the
Schwarzschild orbits, as most studies on the topics of general relativity do.

In this paper, we first present, in Section 2, three explicit analytic
expressions for the orbits in the Schwarzschild geometry: one is for
periodic orbits [6] (including a special case that we call asymptotic), and
two are terminating orbits. The explicit analytic expressions that we derive
not only describe the precise features of the orbits (periodic, precessing,
non-periodic, terminating, etc.) but also clearly indicate two physical
parameters which can be used to characterize these orbits. These two
dimensionless parameters are specific combinations of the following physical
quantities: the total energy and angular momentum of the planet, the masses
of the massive object and the planet, and, of course, the universal
gravitation constant $G$ and the speed of light $c$. These two physical
dimensionless parameters were first used by one of us in ref.6. We shall
refer to these two quantities as the energy eccentricity parameter $e$ and
the gravitational field parameter $s$ respectively (or simply as the energy
parameter and the field parameter). They will be defined in Section 2. We
will use neither the common convention of setting $G=c=1$, nor the energy
and angular momentum of the planet by themselves, as the physical parameters
for characterizing the orbits. With the energy parameter $(0\leq e\leq 1)$
plotted on the horizontal axis and the field parameter $(0\leq s\leq \infty
) $ plotted on the vertical axis, the parameter space for all possible
orbits will be shown to be divisible into three sectors, which we call
Regions I, II and II', that have clearly defined boundaries. Region I
permits periodic and terminating orbits. Regions II and II' terminating
orbits only.

In Section 3, we describe Region I for the orbits in greater detail. We
first divide Region I by lines each of which represents orbits described by
elliptic functions of the same modulus $k$. We then give a more physical
division of Region I which consists of nearly horizontal lines each of which
represents orbits that have the same precession angle $\Delta \phi $, and of
bent vertical lines each of which represents orbits that have the same
"true" eccentricity $\varepsilon $. The terminating orbits will be
characterized by two parameters one of which is the angle at which the
planet enters the center of the blackhole, and the other being the initial
distance of the planet from the star or blackhole. In Sections 4 and 5, we
describe Regions II and II' in which all orbits are terminating, and we
again divide Region II by curves of constant modulus $k$ each of which
describes orbits with the same modulus. Regions II and II' are separated by
the Schwarzschild horizon. Thus the grid of our "map" can be used to
describe all possible orbits in the Schwarzschild geometry in their
entirety. The observational data related to a planet's orbit about some
giant star or blackhole can be directly identified with a point having
certain coordinates $(e,s)$ on our map, which can then be used for
estimating the physical characteristics associated with the star or
blackhole and that of the planet itself, assuming that the star or blackhole
is not spinning very fast. For the Kerr geometry, another dimensionless
quantity, which is clearly the ratio of the spin angular momentum per unit
mass of the blackhole to the orbital angular momentum per unit mass of the
planet, should enter into the consideration. In Section 6, we briefly
discuss a possible extension of our results to the case involving a slowly
spinning blackhole, at least to the first order perturbation, by rescaling
the physical parameters involved.

In Section 7, we study the deflection of light by the gravitational field of
a very massive object. A single dimensionless parameter will be used to
characterize the region. We show that here too, we should divide the region
into three sectors, which we again call Regions I, II and II', and we
present three analytic expressions for the trajectories of light applicable
in these different regions. Region I has trajectories of light that get
deflected, and Regions II and II' have trajectories of light that are
absorbed by and terminate at the blackhole.

In Section 8, we give a summary of our results. Proofs of many interesting
analytic relations among the parameters appearing in these studies are given
in several appendices. Since our results presented in this paper cover
gravitational fields of all ranges, from the weak field produced by the sun
of our solar system, for example, to the very strong field produced by a
blackhole, we want to avoid referring to the massive object that produces
the gravitational field as a blackhole, and prefer to refer to it as the
star or blackhole, and we shall refer to the object of a much smaller mass
that orbits around it as the planet or the particle.

We have supplemented our many analytic results with numerous tables that
present various physical quantities such as the minimum and maximum
distances of the planet from the star and the angles of precession of the
orbits that are calculated from our analytic expressions, as well as
numerous figures that show various kinds of orbits of the planet and various
kinds of deflection of light.

\section{Analytic Expressions for the Orbits}

We consider the Schwarzschild geometry, i.e. the static spherically
symmetric gravitational field in the empty space surrounding some massive
spherical object such as a star or a blackhole of mass $M$. The
Schwarzschild metric for the empty spacetime outside a spherical body in the
spherical coordinates $r,\theta ,\phi $ is [1]

\begin{equation}
dl^{2}=c^{2}\left( 1-\frac{\alpha }{r}\right) dt^{2}-\left( 1-\frac{\alpha }{%
r}\right) ^{-1}dr^{2}-r^{2}d\theta ^{2}-r^{2}\sin ^{2}\theta d\phi ^{2}
\end{equation}

where

\begin{equation}
\alpha =\frac{2GM}{c^{2}}
\end{equation}

is known as the Schwarzschild radius, $G$ is the universal gravitation
constant, and $c$ is the speed of light. If $[x^{\mu }]=(t,r,\theta ,\phi )$%
, then the worldline $x^{\mu }(\tau )$, where $\tau $ is the proper time
along the path, of a particle moving in the equatorial plane $\theta =\pi /2$%
, satisfies the equations [1]

\begin{equation}
\left( 1-\frac{\alpha }{r}\right) \overset{\cdot }{t}=\kappa ,
\end{equation}

\begin{equation}
c^{2}\left( 1-\frac{\alpha }{r}\right) \overset{\cdot }{t}^{2}-\left( 1-%
\frac{\alpha }{r}\right) ^{-1}\overset{\cdot }{r}^{2}-r^{2}\overset{\cdot }{%
\phi }^{2}=c^{2},
\end{equation}

\begin{equation}
r^{2}\overset{\cdot }{\phi }=h,
\end{equation}

where the derivative $\overset{\cdot }{}$ represents $d/d\tau $. The
constant $h$ is identified as the angular momentum per unit mass of the
planet, and the constant $\kappa $ is identified to be

\begin{equation*}
\kappa =\frac{E}{m_{0}c^{2}},
\end{equation*}

where $E$ is the total energy of the planet in its orbit and $m_{0}$ is the
rest mass of the planet at $r=\infty $. Substituting eqs.(3) and (5) into
(4) gives the 'combined' energy equation [1]

\begin{equation}
\overset{\cdot }{r}^{2}+\frac{h^{2}}{r^{2}}\left( 1-\frac{\alpha }{r}\right)
-\frac{c^{2}\alpha }{r}=c^{2}(\kappa ^{2}-1).
\end{equation}

Substituting $dr/d\tau =(dr/d\phi )(d\phi /d\tau )=(h/r^{2})(dr/d\phi )$
into the combined energy equation gives the differential equation for the
orbit of the planet

\begin{equation}
\left( \frac{du}{d\phi }\right) ^{2}=\alpha u^{3}-u^{2}+Bu+C
\end{equation}

where $u=1/r$, \ $B=2GM/h^{2}$, \ $C=c^{2}(\kappa ^{2}-1)/h^{2}$. Following
Whittaker [2], it is convenient to change variable from $u$ to a
dimensionless quantity $U$ defined by

\begin{equation}
U=\frac{1}{4}\left( \frac{\alpha }{r}-\frac{1}{3}\right) =\frac{1}{4}\left(
\alpha u-\frac{1}{3}\right) ,
\end{equation}

or $u=4U/\alpha +1/(3\alpha )$ so that eq.(7) becomes

\begin{equation}
\left( \frac{dU}{d\phi }\right) ^{2}=4U^{3}-g_{2}U-g_{3}
\end{equation}

where

\begin{eqnarray}
g_{2} &=&\frac{1}{12}-s^{2}  \notag \\
g_{3} &=&\frac{1}{216}-\frac{1}{12}s^{2}+\frac{1}{4}(1-e^{2})s^{4},
\end{eqnarray}

and where

\begin{equation}
e=\left[ 1+\frac{h^{2}c^{2}(\kappa ^{2}-1)}{(GM)^{2}}\right] ^{1/2}
\end{equation}

and

\begin{equation}
s=\frac{GM}{hc}.
\end{equation}

The two dimensionless parameters $e$ and $s$, which are defined by the two
above equations and which we call the energy and field parameters
respectively, will be the principal parameters we shall use for
characterizing the orbit of a planet. It will be noted that the constant $%
c^{2}(\kappa ^{2}-1)$ which is $<0$ for a bound orbit, can be identified
with $2E_{0}/m$ in the Newtonian limit, where $E_{0}$ is the sum of the
kinetic and potential energies and is $<0$ for a bound orbit, and $m$ is the
mass of the planet (which approaches $m_{0}$), and that

\begin{equation}
e\simeq \left[ 1+\frac{2E_{0}h^{2}}{m(GM)^{2}}\right] ^{1/2}
\end{equation}

is the "eccentricity" of the orbit. Also, in the small $s$ limit, the orbit
equation can be shown to be given by

\begin{equation}
\frac{1}{r}\simeq \frac{GM}{h^{2}}[1-e\cos (1-\delta )\phi ],
\end{equation}

where $\delta \simeq 3(GM)^{2}/(hc)^{2}$. Thus $r$ assumes the same value
when $\phi $ increases to $\phi +2\pi /(1-\delta )$. Comparing this with the
increase of $\phi $ from $\phi $ to $\phi +2\pi $, the ellipse will rotate
about the focus by an amount which is the angle of precession

\begin{equation}
\Delta \phi \simeq \frac{2\pi }{1-\delta }-2\pi \simeq 2\pi \delta =\frac{%
6\pi (GM)^{2}}{h^{2}c^{2}}.
\end{equation}

This is the well known approximate expression for the precession angle for
the case of very small $s$. The limiting case for $\delta =0$ is the well
known orbit equation in Newtonian mechanics. We should note that while the
limit $s=0$ (and thus $\delta =0$) cannot be strictly correct in principle
so long as $M\neq 0$, this limit can be used for many practical cases with
great accuracy as evidenced by the predictions of Newtonian mechanics. A
special case of these Newtonian orbits is the circular orbit of radius $%
r=h^{2}/GM$ for $e=0$.

We now derive the exact analytic solutions of eq.(9) and classify the three
possible solutions from a purely mathematical viewpoint, and we shall
consider their physical interpretations in the next section. We first define
the discriminant $\Delta $ of the cubic equation

\begin{equation}
4U^{3}-g_{2}U-g_{3}=0
\end{equation}

by

\begin{equation}
\Delta =27g_{3}^{2}-g_{2}^{3}.
\end{equation}

The three roots of the cubic equation (16) are all real for the case $\Delta
\leq 0$. We call the three roots $e_{1},e_{2},e_{3}$ and arrange them so
that $e_{1}>e_{2}>e_{3}$; the special cases when two of the roots are equal
will be considered also. For the case $\Delta >0$, the cubic equation (16)
has one real root and two roots that are complex conjugates. The analytic
solutions of eq.(9) that we shall present will give the distance $r$ of the
planet from the star or blackhole in terms of the Jacobian elliptic
functions that have the polar angle $\phi $ in their argument and that are
associated with a modulus $k$ that will be defined. Instead of writing $r$,
we use the dimensionless distance $q$ measured in units of the Schwarzschild
radius $\alpha $ and defined by

\begin{equation}
q\equiv \frac{r}{\alpha }\equiv \frac{1}{\alpha u}.
\end{equation}

The dimensionless distance $q$ is related to $U$ of eq.(8) by

\begin{equation}
\frac{1}{q}=\frac{1}{3}+4U.
\end{equation}

We now give the three analytic solutions of eq.(9) in the following.

Solution (A1) For $\Delta \leq 0$, $e_{1}>e_{2}\geq U>e_{3}$.

Writing the right-hand side of eq.(9) as $4(e_{1}-U)(e_{2}-U)(U-e_{3})$,
eq.(9) can be integrated with $\phi $ expressed in terms of the inverse
Jacobian $sn$ function [7]. After a little algebra and some rearrangement,
the equation for the orbit is found to be

\begin{eqnarray}
\frac{1}{q} &=&\frac{1}{3}+4e_{3}+4(e_{2}-e_{3})sn^{2}(\gamma \phi ,k) 
\notag \\
&=&\frac{1}{3}+4e_{3}+4(e_{2}-e_{3})\frac{1-cn(2\gamma \phi ,k)}{%
1+dn(2\gamma \phi ,k)},
\end{eqnarray}

where the point at $\phi =0$ has been chosen to give $U=e_{3}$. The constant 
$\gamma $ appearing in the argument, and the modulus $k$, of the Jacobian
elliptic functions are given in terms of the three roots of the cubic
equation (16) by

\begin{eqnarray}
\gamma &=&(e_{1}-e_{3})^{1/2}, \\
k^{2} &=&\frac{e_{2}-e_{3}}{e_{1}-e_{3}}.
\end{eqnarray}

where $e_{1},e_{2},e_{3}$ are given by

\begin{eqnarray}
e_{1} &=&2\left( \frac{g_{2}}{12}\right) ^{1/2}\cos \left( \frac{\theta }{3}%
\right) ,  \notag \\
e_{2} &=&2\left( \frac{g_{2}}{12}\right) ^{1/2}\cos \left( \frac{\theta }{3}+%
\frac{4\pi }{3}\right) ,  \notag \\
e_{3} &=&2\left( \frac{g_{2}}{12}\right) ^{1/2}\cos \left( \frac{\theta }{3}+%
\frac{2\pi }{3}\right) ,
\end{eqnarray}

and where

\begin{equation}
\cos \theta =g_{3}\left( \frac{27}{g_{2}^{3}}\right) ^{1/2}.
\end{equation}

Equation (20) was first given in ref.6 using a slightly different approach
that was initiated by Whittaker [2]. In addition, eq.(20) was shown to
reduce to eq.(14) for the case of very small $s$ which in turn gave the
known approximate precession angle given by eq.(15). The modulus $k$ of the
elliptic functions has a range $0\leq k^{2}\leq 1$. Since the elliptic
functions $sn$, $cn$ and $dn$ are all periodic functions of $\phi $ for $%
0\leq k^{2}<1$, we shall call this solution for the orbit the periodic
solution. For the special case of $k^{2}=1$, since $sn(\gamma \phi ,1)=\tanh
(\gamma \phi )$, $cn(\gamma \phi ,1)=dn(\gamma \phi ,1)=\sec h(\gamma \phi )$%
, the solution is no longer periodic, and we shall refer to it as the
asymptotic periodic solution.

The period of $cn(2\gamma \phi ,k)$ is $4K(k)$, and the period of $%
dn(2\gamma \phi ,k)$ and of $sn^{2}(\gamma \phi ,k)$ is $2K(k)$, where $K(k)$
is the complete elliptic integral of the first kind [7]. For $k=0$, $%
sn(x,0)=\sin x,$ $cn(x,0)=\cos x,$ $dn(x,0)=1$. As $k^{2}$ increases from $0$
to $1$, $K(k)$ increases from $\pi /2$ to $\infty $. The distance $r$ of the
planet from the center of the star or blackhole assumes the same value when
its polar angle $\phi $ increases from $\phi $ to $\phi +4K/(2\gamma )=\phi
+2K/\gamma $. Comparing this with the increase of $\phi $ from $\phi $ to $%
\phi +2\pi $ in one revolution, the perihelion (or the aphelion) will rotate
by an amount

\begin{equation}
\Delta \phi =\frac{2K(k)}{\gamma }-2\pi ,
\end{equation}

which is the exact expression for the precession angle. For $k^{2}$ close to
the value $1$, the planet can make many revolutions around the star or
blackhole before assuming a distance equal to its initial distance. Thus if $%
n$ is the largest integer for which $2K(k)/\gamma $ is equal to or greater
than $2n\pi $, the angle of precession should be more appropriately defined
as $2K(k)/\gamma -2n\pi $. For the sake of consistency, however, we shall
stick to the definition given by eq.(25).

For the case of very small $s$ and to the order of $s^{2}$, it was shown in
ref.6 that $\gamma \simeq \lbrack 1-(3-e)s^{2}]/2$, $k^{2}\simeq 4es^{2}$, $%
K(k)\simeq \pi (1+es^{2})/2$, and substituting these into eq.(25) gives the
well known approximate result given by eq.(15).

For these periodic orbits, we note that the maximum distance $r_{\max }$
(the aphelion) of the planet from the star or blackhole and the minimum
distance $r_{\min }$ (the perihelion) of the planet from the star or
blackhole, or their corresponding dimensionless forms $q_{\max }$ $(=r_{\max
}/\alpha )$ and $q_{\min }$ $(=r_{\min }/\alpha ),$ are obtained from
eq.(20) when $\gamma \phi =0$ and when $\gamma \phi =K(k)$ respectively, and
they are given by

\begin{equation}
\frac{1}{q_{\max }}=\frac{1}{3}+4e_{3},
\end{equation}

and

\begin{equation}
\frac{1}{q_{\min }}=\frac{1}{3}+4e_{2},
\end{equation}

where $e_{2}$ and $e_{3}$ are determined from eqs.(23), (24) and (10) in
terms of $e$ and $s$.

Although we call the orbits given by this solution for $0\leq k^{2}<1$
periodic, they are not necessarily closed orbits. It is seen from the
precession discussed above that for $\Delta \phi =f\pi $, unless $f$ is a
rational number, the orbit will not close and it is not strictly a closed
periodic orbit. However, for all practical purposes, any irrational number
when truncated becomes a rational number, and thus the orbit will be closed.
The distinction of closed and non-closed orbits depending on whether $f$ is
rational or irrational is of course of profound theoretical interest [5].

For a general periodic orbit that precesses, the general or true
eccentricity $\varepsilon $ of the orbit is defined by

\begin{equation}
\varepsilon \equiv \frac{r_{\max }-r_{\min }}{r_{\max }+r_{\min }}=\frac{%
q_{\max }-q_{\min }}{q_{\max }+q_{\min }},
\end{equation}

where $q_{\max }$ and $q_{\min }$ are given by eqs.(26) and (27).

We shall show in the following section that the true eccentricity $%
\varepsilon $ is in general not equal to the energy eccentricity parameter $%
e $ defined by eq.(11), but that $\varepsilon \rightarrow e$ in the limit of 
$s\rightarrow 0$, i.e. in the Newtonian limit. For the special case of $%
\varepsilon =1$ however, we shall show that it coincides with the special
case of $e=1$ for all applicable values of $s$, and that it signifies an
unbounded orbit.

\bigskip

We now proceed to present the second solution.

Solution (A2) For $\Delta \leq 0$, $U>e_{1}>e_{2}>e_{3}$.

We write the right-hand side of eq.(9) as $4(U-e_{1})(U-e_{2})(U-e_{3})$ and
eq.(9) can be integrated with $\phi $ expressed in term of the inverse
Jacobian $sn$ function [7].\ The equation for the orbit is found to be

\begin{equation}
\frac{1}{q}=\frac{1}{3}+4\frac{e_{1}-e_{2}sn^{2}(\gamma \phi ,k)}{%
cn^{2}(\gamma \phi ,k)},
\end{equation}

where $\gamma $, $k$, $e_{1}$, $e_{2}$ and $e_{3}$ are given by
eqs.(21)-(24) as in the first solution. This solution gives a terminating
orbit. The point at $\phi =0$ has been chosen to be given by

\begin{equation}
\frac{1}{q_{1}}=\frac{1}{3}+4e_{1}.
\end{equation}

The planet, starting from the polar angle $\phi =0$ at a distance $q_{1}$
from the blackhole, plunges into the center of the blackhole when its polar
angle $\phi _{1}$ is given by $cn(\gamma \phi _{1},k)=0$, i.e. when

\begin{equation*}
\phi _{1}=\frac{K(k)}{\gamma },
\end{equation*}

where $\gamma $ and $k$ are given by eqs.(21) and (22).

The region of $(e,s)$ where orbits given by solutions A1 and A2 are
applicable will be called Region I, and it will be described in greater
detail in Section 3. Thus each point $(e,s)$ of parameter space in Region I
represents two distinct orbits, one periodic and one terminating. At the
same coordinate point, the characteristic quantities that describe the two
distinct orbits are related. For example, by noting $e_{1}+e_{2}+e_{3}=0$
and from eqs.(26) and (27), $q_{1}$ can be expressed as

\begin{equation}
\frac{1}{q_{1}}=1-\left( \frac{1}{q_{\min }}+\frac{1}{q_{\max }}\right) ,
\end{equation}

where $q_{\min }$ and $q_{\max }$ are the minimum and maximum distances for
the periodic orbit at the same coordinate points $(e,s)$. It will be noted
that $q_{1}$ is less than $q_{\min }$, i.e. for the terminating orbit the
planet is assumed initially to be closer to the blackhole than the $q_{\min
} $ for the associated periodic orbit, except at $k^{2}=1$ where $%
q_{1}=q_{\min }$ and the planet has a circular instead of a terminating
orbit that will be explained later.

We note that the terminating orbit equation (29) presented has no
singularity at the Schwarzschild horizon $q=1$, because, as is well known, $%
q=1$ is a coordinate singularity and not a physical singularity. The orbit
obtained from continuing $\phi $ beyond the value $\phi _{1}=K(k)/\gamma $
may become interesting if the concept of "whitehole" turns out to be of
physical relevance.

For now, the orbits in Region I are characterized mathematically by $\Delta
\leq 0$.

We now present the third solution.

Solution (B) For $\Delta >0$.

Define

\begin{eqnarray}
A &=&\frac{1}{2}\left( g_{3}+\sqrt{\frac{\Delta }{27}}\right) ^{1/3},  \notag
\\
B &=&\frac{1}{2}\left( g_{3}-\sqrt{\frac{\Delta }{27}}\right) ^{1/3},
\end{eqnarray}

where $g_{3}$ and $\Delta $ are defined by eqs.(10) and (17). The real root
of the cubic equation (16) is given by

\begin{equation}
a=A+B
\end{equation}

and the two complex conjugate roots $b$ and $\overline{b}$ are $-(A+B)/2\pm
(A-B)\sqrt{3}i/2$. We further define

\begin{equation}
\gamma =[3(A^{2}+AB+B^{2})]^{1/4}
\end{equation}

and

\begin{equation}
k^{2}=\frac{1}{2}-\frac{3(A+B)}{4\sqrt{3(A^{2}+AB+B^{2})}}=\frac{1}{2}-\frac{%
3a}{4\gamma ^{2}}.
\end{equation}

Writing the right-hand side of eq.(9), with $U\geq a$, as $4(U-a)(U-b)(U-%
\overline{b})$, eq.(9) can be integrated with $\phi $ expressed in terms of
the inverse Jacobian $cn$ function [7].\ We find the equation for the orbit
to be

\begin{equation}
\frac{1}{q}=\frac{1}{3}+4\frac{\gamma ^{2}+a-(\gamma ^{2}-a)cn(2\gamma \phi
,k)}{1+cn(2\gamma \phi ,k)}.
\end{equation}

This is a terminating orbit. The initial distance $q_{2}$ of the planet at $%
\phi =0$ has been chosen to be given by

\begin{equation}
\frac{1}{q_{2}}=\frac{1}{3}+4a.
\end{equation}

It plunges into the center of the blackhole when its polar angle $\phi =\phi
_{2}$ is given by

\begin{equation*}
\phi _{2}=\frac{K(k)}{\gamma },
\end{equation*}

where $\gamma $ and $k$ are given by eqs.(34) and (35). Again, we note that
the orbit equation (36) has no singularity at $q=1$.

The region of $(e,s)$ where orbits given by eq.(36) are applicable will be
divided into two sectors called Regions II and II', the boundary between
which will be defined later. They have terminating orbits only. For now, the
orbits in Regions II and II' are characterized mathematically by $\Delta >0$.

As for the initial points of the orbits discussed above, by comparing
eq.(19) with the orbit equations (20), (29) and (36), and with eqs.(26),
(30) and (37), we already noted that our choice of $\phi =0$ in our orbit
equations is such that it gives $U=e_{3}$, $e_{1}$, and $a$ respectively
that in turn give $q=q_{\max }$, $q_{1}$, and $q_{2}$ as the initial
distances of the planet from the star or blackhole. We then note from eq.(9)
that $dU/d\phi =0$ and hence $dr/d\phi =0$ for the planet at these initial
points of the trajectories, i.e. the trajectory or more precisely the
tangent to the trajectory at $\phi =0$ is perpendicular to the line joining
the planet to the star or blackhole. All this will be seen in the figures
presented later, and all our references to the initial position of the
planet from here onward assume that the trajectory (as $\phi $ increases
from $0$) of the planet at its initial position is perpendicular to the line
joining the planet to the star or blackhole.

\section{Region I}

Consider the orbits expressed by eqs.(20) and (29) given by solutions A1 and
A2 and characterized mathematically by $\Delta \leq 0$. We call the region
covered by the associated range of values for $(e,s)$ Region I.

To gain a preliminary perspective, consider the earth (as the planet) and
the sun (as the star) in our solar system. Substituting the mass of the sun $%
M=M_{S}=1.99\times 10^{30}kg$ and the angular momentum of the earth per unit
mass of the earth $h=4.48\times 10^{15}m^{2}/s$, we find $s=0.983\times
10^{-4}$. The energy eccentricity parameter $e$, which is equal to the true
eccentricity $\varepsilon $ of the earth's orbit for such a very small $s$
value, is known to be about $0.017$. The approximate relation $k^{2}\simeq
4es^{2}$ gives the squared modulus of the elliptic functions that describe
the earth's orbit to be $k^{2}=0.657\times 10^{-9}$. We see that for the
planetary system that is familiar to us, the values of $s$ and $k^{2}$ are
very small indeed. We may also note that the Schwarzschild radius $\alpha
=2GM_{S}/c^{2}\simeq 3$ $km$ would be well inside the sun which has a radius
of $6.4\times 10^{3}km$. The earth's dimensionless distance is $q\simeq
5\times 10^{7}$ from the sun's center. For this value of $s$, with $q_{\min
}\simeq q_{\max }\simeq 5\times 10^{7}$, the orbit given by eq.(29) from
solution A2 would require the initial position $q_{1}$ of a planet to be $%
\simeq 1$ according to eq.(31), i.e. the planet would have to be at a
distance equal to the Schwarzschild radius from the center of the sun for it
to have a terminating orbit which plunges to the center of the sun.
Therefore the terminating orbit given by eq.(29) is inapplicable for our
solar system. The periodic orbits, on the other hand, are perfectly valid.

However, for cases when the massive object is a gigantic mass concentrated
in a small radius such as a blackhole, all the possibilities presented here
may arise. As the field parameter $s$ increases from $0$, the modulus $k$ of
the elliptic functions that describe the planet's orbits also increases.
From eqs.(21)-(24), it is seen that several steps are needed to relate $%
k^{2} $ to $e$ and $s$. In Appendix A, we show that a direct relationship
between $k^{2}$ and $e$ and $s$ can be established, and it is given by

\begin{eqnarray}
\frac{1-18s^{2}+54(1-e^{2})s^{4}}{(1-12s^{2})^{3/2}} &=&\frac{%
(2-k^{2})(1+k^{2})(1-2k^{2})}{2(1-k^{2}+k^{4})^{3/2}} \\
&=&\cos \theta .
\end{eqnarray}

The $\cos \theta $ of eq.(39) is the same $\cos \theta $ that appears in
eq.(24), and, in particular, it is equal to $1,0,-1$ for $k^{2}=0,1/2,1$
respectively.

The curve represented by $k^{2}=1$, after setting $\cos \theta =-1$ in
eq.(39), can be readily shown to give a quadratic equation $%
27(1-e^{2})^{2}s^{4}-2(1-9e^{2})s^{2}-e^{2}=0$ that gives

\begin{equation}
s_{1}^{2}=\frac{1-9e^{2}+\sqrt{(1-9e^{2})^{2}+27e^{2}(1-e^{2})^{2}}}{%
27(1-e^{2})^{2}},
\end{equation}

for $0\leq e<1$, and $s_{1}^{2}=1/16$ for $e=1$. Equation (40) representing $%
k^{2}=1$ gives the upper boundary (for the values of $s$) of Region I (the
uppermost heavy solid line in Fig.1); it extends from $s_{1}=\sqrt{2/27}%
=0.272166$ for $e=0$ to $s_{1}=1/4=0.250000$ for $e=1$, i.e. a line that is
nearly parallel to the $e$-axis. Thus Region I is a region bounded by $0\leq
e\leq 1$, and by $0\leq s\leq s_{1}$ where $s_{1}$ is given by eq.(40), in
which the squared modulus of the elliptic functions that describe the orbits
cover the entire range $0\leq k^{2}\leq 1$.

We now use eq.(38) to give a plot of lines of constant $%
k^{2}=0.001,0.01,0.1,0.3,...,1$ as shown in Fig.1. These lines conveniently
divide Region I into regions of increasing field strengths as $k^{2}$
increases from $0$ to $1$. On a point representing a particular $k^{2}$ and
a particular $e$ value, $s$ can be determined from eq.(38) and the orbit is
then given by eq.(20) using eqs.(98), (10) and (21). The values of $s$ on
these constant $k^{2}$ lines for the values of $e=0.1,0.2,...,1.0$ are given
in Table 1 which thus give the coordinates $(e,s)$ of the points on the
lines representing different values of $k^{2}$. These coordinate points $%
(e,s)$ from Table 1 are used to give the following tables: Tables 2 and 3
give the values of $q_{\max }$ and $q_{\min }$ for the orbits obtained from
eqs.(26) and (27). Note that the dimensionless distance $q$ is in units of
the Schwarzschild radius $\alpha $ which depends on the mass $M$ of the star
or blackhole corresponding to that particular coordinate point, and thus one
should not compare $q$ at two different coordinate points just by their
absolute values alone. Table 4 presents the values of the precession angle
in units of $\pi $, i.e. $\Delta \phi /\pi $, obtained from eq.(25). Table 5
presents the values of the true eccentricity $\varepsilon $ obtained from
eq.(28). Tables 2-5 are to be used in conjunction with Table 1 for
identifying the locations $(e,s)$ of the corresponding quantities that are
presented. The physical quantities presented in Tables 2-5 together with the
coordinates $(e,s)$ given in Table 1 now give all possible periodic orbits
in the Schwarzschild geometry in its entirety. That is, the coordinates $%
(e,s)$ of a planet orbiting a non-spinning blackhole can be identified if
the observation data on $r_{\min },r_{\max },\varepsilon $ and $\Delta \phi $
can be collected. Region I shown in Fig.1 is where orbits given by eqs.(20)
and (29) apply. In Sections 4 and 5, we shall discuss Regions II and II'
which are shown above Region I in Fig.2 where orbits given by eq.(36) apply.
As an example of application of Tables 1-5, from the second row and second
column of Tables 1-5 and using only two significant figures, for orbits with 
$e=0.10,s=0.11,k^{2}=0.010,$ we find from Tables 2-5 that $q_{\max
}=50,q_{\min }=34,\Delta \phi /\pi =0.079$ or $\Delta \phi =14^{\circ },$
and $\varepsilon =0.19$, i.e. orbits with those seemingly small values of $s$
and $k^{2}$ give a precession angle of $14^{\circ }$ per revolution that is
already very large compared to those encountered in our solar system for
which the precession angle is only $3.8"$ per century for the earth's orbit
(for which $s\simeq 0.983\times 10^{-4},k^{2}\simeq 0.657\times
10^{-9},\varepsilon \simeq e\simeq 0.017$), and the value of the true
eccentricity $\varepsilon $ of these orbits is already quite different from
their energy parameter $e$. We thus appreciate that the range of values for $%
s$ given by $0\leq s\leq s_{1}$ for Region I, where $s_{1}$ ranges from $%
0.276166$ for $e=0$ to $0.25$ for $e=1$, is not as small as it seems (noting
also that $0\leq k^{2}\leq 1$), and that the classical Newtonian orbits are
restricted to a very narrow strip of the region indeed for which $s\simeq 0$
and $k^{2}\simeq 0,$ and for which $\varepsilon \simeq e$ for $0\leq e\leq 1$%
.

Although the lines of constant $k^{2}$ in Region I conveniently associate
the orbits with the orbit equations for the periodic and terminating orbits
given by eqs.(20) and (29) and with the physical parameters given in Tables
2-5, the precession angle $\Delta \phi $ and the true eccentricity $%
\varepsilon $ are more physically meaningful parameters that can be
associated with the description of the orbit. The expressions for $\Delta
\phi $ and $\varepsilon $ in terms of $k$ and $s$ are given by eq.(99) in
Appendix A and eq.(100) in Appendix B. For a given value of $\Delta \phi $
and of $e$, we can use eqs.(99) and (38) to solve for $s$ (and $k$) (using a
numerical program such as FSOLVE in MAPLE) and thus locate its coordinate $%
(e,s)$; and similarly for a given value of $\varepsilon $ and of $e$, we can
use eqs.(100) and (38) to solve for $s$ (and $k$). The relationship of $e$
and $s$ with $\varepsilon $ is simpler for $k^{2}=1$ and will be discussed
later (see eqs.(49) and (50)). In Fig.3, we present lines of constant $%
\Delta \phi /\pi $ (that are nearly horizontal) and lines of constant $%
\varepsilon $ (that are bent vertical) in Region I, and the corresponding
tables for their coordinates are presented in Tables 6 and 7. We note that
because $\Delta \phi $ given by eq.(25) depends on $\gamma $ given by
eq.(21) as well as on $K(k)$, the line of constant $\Delta \phi $ does not
coincide with the line of constant $k^{2}$ except for $k^{2}=1$. We note
also that the line of constant $\varepsilon $ does not coincide with the
(vertical) line of constant $e$ except for $\varepsilon =e=1$. We show in
Appendix B that it is only for a very thin strip of region, where $s$ is
between zero and some very small positive value, that $\varepsilon \simeq e$
which applies in the Newtonian limit. We also show in Appendix B that $%
\varepsilon =e$ when $e=1$ exactly. The distinction between $e$ defined by
eq.(11) or eq.(13) with $\varepsilon $ defined by eq.(28) in the Newtonian
or non-Newtonian theory has never been clearly recognized previously.

With Fig.3 which has curves of constant $\Delta \phi /\pi $ and constant $%
\varepsilon $ in place, Region I is now partitioned into cells with the
coordinate points specified by $(\Delta \phi /\pi ,\varepsilon )$. We have a
clear idea what the orbits of a planet would be like at points within each
cell in terms of their precession angle and true eccentricity, and the
coordinates of these orbits $(e,s)$ then give the energy and field
parameters corresponding to these orbits. In Fig.4, we present examples of
periodic and unbounded orbits, plotted in polar coordinates $(q,\phi )$,
corresponding to various precession angles of $\pi /6,\pi /3,\pi /2,\pi
,3\pi /2,2\pi ,\infty $ (vertically from top to bottom) for values of $%
e=0,0.5,1$ (horizontally from left to right), where the star or blackhole is
located at the origin. We first note that the orbits for which $e<1$ are
periodic and closed because $f$ is a rational number in $\Delta \phi =f\pi $
for each one of them. The precession angle can be seen from the heavy solid
line that marks the trajectory (as $\phi $ increases) from the initial point
at $\phi =0$ to the first point at which the distance from the origin is
equal to the distance at $\phi =0$. The true eccentricity of the orbits is $%
\varepsilon $ given by eq.(28). For example, for the orbit of Fig.4 (a1) for 
$\Delta \phi =\pi /6$, $e=0$, $\varepsilon $ is far from zero which can be
seen from the $q_{\min }$ and $q_{\max }$ in the figure, and it can be more
accurately calculated to be equal to $0.22629$. For each of the unbounded
orbits characterized by $e=1$, the incoming trajectory coming from infinity
at $\phi =0$ makes an angle $\phi $ with the outgoing trajectory going to
infinity given by $\phi =2K(k)/\gamma =2\pi +\Delta \phi $ from eq.(25), as
can be seen in some of the figures presented. The case $\Delta \phi /\pi
=\infty $ corresponding to the special case of $k^{2}=1$ will be discussed
later in this section for which the planet starting from $q_{\max }$ ends up
circling the blackhole with a radius that approaches $q_{\min }$ (see
Fig.6d).

Generally, if we are given a coordinate point in Fig.3, for example, a point
on $e=0.5$ just above the $\Delta \phi /\pi =1/3$ line slightly to the left
of the $\varepsilon =0.6$ curve (where $\varepsilon =0.581431...$ and $%
s=0.194229...$), then we find $\Delta \phi =60.4706...$ degrees or $\Delta
\phi /\pi =0.33594...$, and part of the orbit is shown in Fig.5. Whether the
orbit will close on itself depends on whether $\Delta \phi /\pi $ is or is
not a rational number in principle, even though, as we mentioned before, a
truncated number in practice is always a rational number and the orbit will
be a closed one. We only show part of the orbit in Fig.5 as the subsequent
path is clear from the angle of precession and true eccentricity of the
orbit and we are not concerned with how many "leaves" the orbit is going to
create. Figure 3 (or one with even more curves of constant $\Delta \phi /\pi 
$ and constant $\varepsilon $) is a very useful map that can be used
fruitfully with any observation data that are obtained for any planet.

Besides the special case $k^{2}=1$, the case of $k^{2}=1/2$ is also somewhat
special in that it allows many relationships to be expressed simply and
explicitly. We present some of these simple relations for $k^{2}=1/2$ in
Appendix C. \ It is to be noted from Fig.1 that the line of constant $%
k^{2}=1/2$ is very close to the boundary given by $k^{2}=1$. The line of
constant $k^{2}=1/2$ for Region II, on the other hand, is closer to dividing
the region approximately into two halves, as shown in Fig.2. The $k^{2}=1/2$
curve for Region II will be discussed in Section 4.

The terminating orbits in Region I given by eq.(29) can be characterized by
the planet's initial position $q_{1}$ given by eq.(31), and by the angle $%
\phi _{1}$ at which the planet enters the center of the blackhole. It is
interesting to note that even for these terminating orbits, the precession
angle still has an "extended" meaning and use that we shall describe. It is
clear from eq.(29) that the orbit terminates, i.e. $q$ becomes zero when $%
\gamma \phi _{1}=K(k)$, but if the orbit is continued (by continuing to
increase $\phi $), $q$ would assume its initial value at $\phi =0$ when $%
\gamma \phi ^{\prime }=2K(k)$, producing a "precession angle" of $\Delta
\phi =\phi ^{\prime }-2\pi =2K(k)/\gamma -2\pi $ which is equal to the
precession angle for the corresponding periodic orbit at the same coordinate
point $(e,s)$. Since $\phi ^{\prime }=2\phi _{1}$, the polar angle $\phi
_{1} $ at which the path of the terminating orbit enters the center of the
blackhole is related simply to the precession angle of the periodic orbit by 
$\phi _{1}=\Delta \phi /2+\pi $, or

\begin{equation*}
\frac{\phi _{1}}{\pi }=\frac{1}{2}\frac{\Delta \phi }{\pi }+1.
\end{equation*}

As $\phi _{1}/\pi $ can be easily calculated from $\Delta \phi /\pi $ for
the periodic orbits given in Table 4, we do not tabulate it separately. The
values of $q_{1}$ are presented in Table 8, and we note the small range $%
1\leq q_{1}\leq 2.25$ for the entire Region I. Examples of these terminating
orbits are presented in Fig.6. The dotted line represents the continuation
of the orbit when $\phi $ is continued beyond $\phi _{1}$.

Before we discuss Regions II and II', we want to describe three special
cases: the case of $k^{2}=0$ which, as we shall see, is not of any interest
but must be included for completeness; the case of $k^{2}=1$ which gives the
upper boundary of Region I (and lower boundary of Region II); and the case
of $e=1$ which gives the right boundary of Region I (and of Regions II and
II') (see Figs.1 and 2).

(i) The Special Case of $k^{2}=0$

The line of $k^{2}=0$ coincides with the $s=0$ axis in Fig.1. To show this,
we note that $k^{2}=0$ implies $\theta =0$ from eq.(95). Substituting $%
\theta =0$ into eq.(24) gives $s=0$ when we use the expressions in eq.(10)
for $g_{2}$ and $g_{3}$. The we find $g_{2}=1/12$ and $g_{3}=1/216$, and
from eq.(98), we find

\begin{eqnarray*}
e_{1} &=&\frac{1}{6}, \\
e_{2} &=&e_{3}=-\frac{1}{12}.
\end{eqnarray*}

Equation (20) then gives $1/q=0$ or $q=\infty $, i.e. it is the limiting
case of zero gravitational field. As we pointed out earlier, the classical
Newtonian case is given by only a very narrow strip represented by $%
k^{2}\simeq 0$ and $s\simeq 0$ for which $q$ is large but finite.

(ii) The Special Case of $k^{2}=1$

It follows from eqs.(38) and (39) that on the line of $k^{2}=1$, $\cos
\theta =-1$. Thus from eqs.(24) and (17), we have

\begin{equation}
\Delta =0
\end{equation}

which can be identified as the "boundary" between Solutions A and B in
Section 2. The range of $s$ values for $\Delta =0$ is $0.25\leq s\leq
0.272166$ for $1\geq e\geq 0$ (see the discussion below eq.(40)), and for
that range of $s$ values, $s\leq 1/2\sqrt{3}=0.288675$ or $s^{2}\leq 1/12$
and therefore $g_{2}>0$ (see eq.(10)). From eq.(41), the relation between $%
g_{2}$ and $g_{3}$ can be more precisely expressed as

\begin{equation*}
\sqrt[3]{g_{3}}=-\sqrt{\frac{g_{2}}{3}}
\end{equation*}

after noting that $g_{3}$ is negative and $g_{2}$ is positive for the values
of $s$ along the line $k^{2}=1$. Also from eq.(98), we note that

\begin{eqnarray}
e_{1} &=&e_{2}=\sqrt{\frac{g_{2}}{12}},  \notag \\
e_{3} &=&-\sqrt{\frac{g_{2}}{3}}.
\end{eqnarray}

The periodic orbits given by eq.(20) become

\begin{equation}
\frac{1}{q}=\frac{1}{3}+2\sqrt{\frac{g_{2}}{3}}\frac{1-5\sec h(2\gamma \phi )%
}{1+\sec h(2\gamma \phi )},
\end{equation}

where

\begin{equation}
\gamma =\left( \frac{3g_{2}}{4}\right) ^{1/4}
\end{equation}

and where the values of $g_{2}$ (and $g_{3}$) are those given by the values
of $e$ and $s$ on the line $k^{2}=1$ that are obtained from eq.(40). The
orbit is not a periodic orbit; it is what we call an asymptotic periodic
orbit. The planet starts from an initial position $q_{\max }$ at $\phi =0$
given by

\begin{equation}
\frac{1}{q_{\max }}=\frac{1}{3}+4e_{3}=\frac{1}{3}-4\sqrt{\frac{g_{2}}{3}}
\end{equation}

and ends up at $\phi =\infty $ circling the star or blackhole with a radius
that asymptotically approaches $q_{\min }$ given by

\begin{equation}
\frac{1}{q_{\min }}=\frac{1}{3}+4e_{2}=\frac{1}{3}+2\sqrt{\frac{g_{2}}{3}}.
\end{equation}

Equations (43)-(46) are explicit and simple equations that give the orbit
equation, $q_{\max }$, and $q_{\min }$ for $k^{2}=1$. In particular, it is
seen from Table 3, for example, that $q_{\min }$ ranges from $2$ for $e=0$
to $9/4=2.25$ for $e=1$, i.e. $q_{\min }$ is still no less than twice the
Schwarzschild radius for the strongest gravitational field that permits the
periodic orbits. However, it is a very small number indeed compared to, say, 
$q_{\min }\simeq 5\times 10^{7}$ for the earth's orbit around the sun.

On this upper boundary $k^{2}=1$ of Region I, the terminating orbit given by
eq.(29) from Solution A2 becomes a circular orbit with a radius $q_{c}$ $%
=q_{1}$, where $q_{1}$ is the initial distance of the planet from the star
or blackhole given by eq.(30). From eqs.(30) and (31) and noting that $%
e_{1}=e_{2}$ for $k^{2}=1$, we find that

\begin{equation}
q_{c}=q_{1}=q_{\min }
\end{equation}

given by eq.(46) (see Tables 3 and 8 for $k^{2}=1$). We shall refer to the
orbits given by eqs.(43) and (47) as the asymptotic periodic and the
asymptotic terminating orbits respectively of Region I. Thus the special
cases given by eqs.(43) and (47) for $k^{2}=1$ of the periodic and
terminating orbits given by eqs.(20) and (29) for solutions A1 and A2
respectively clearly exhibit completely different behaviors from their
counterparts for $0\leq k^{2}<1$. Examples of asymptotic periodic orbits are
shown in Fig.4g. Asymptotic terminating orbits are simply circles of radius
equal to $q_{1}$, as shown in Fig.6(d).

Using eqs.(28), (42), (45) and (46), for $k^{2}=1$ the true eccentricity $%
\varepsilon $ can be shown to be expressible in terms of $g_{2}$ by

\begin{equation}
\varepsilon =\frac{9\sqrt{g_{2}/3}}{1-3\sqrt{g_{2}/3}},
\end{equation}

which can be solved to give $s$ in terms of $\varepsilon $, and then $e$ in
terms of $\varepsilon $ using eq.(38). We find that the coordinates $(e,s)$
of a given $0.6\leq \varepsilon \leq 1$ on the line $k^{2}=1$ are given by

\begin{equation}
e=\frac{\sqrt{(1+\varepsilon )(-3+5\varepsilon )}}{(3-\varepsilon )},
\end{equation}

and

\begin{equation}
s=\frac{\sqrt{(3-\varepsilon )(1+\varepsilon )}}{2(3+\varepsilon )}.
\end{equation}

It is interesting that eqs.(49) and (50) can be used in place of eq.(40) as
parametric equations for determining the coordinates $(e,s)$ of the line $%
k^{2}=1$ as $\varepsilon $ takes the values from $0.6$ to $1$. In
particular, eqs.(49) and (50) allow us to see that the $\varepsilon =const.$
curves are not vertical (except for $\varepsilon =e=1$), and they intersect
the upper boundary $s_{1}$ of Region I for $0.6\leq \varepsilon \leq 1$ (see
Fig.3). The $\varepsilon =0.6$ curve, the boundary curve $s_{1}$, and the $s$%
-axis are concurrent at $e=0$, $s=\sqrt{2/27}$. We can conclude that
periodic orbits with $e=0$ have true eccentricity in the range $0\leq
\varepsilon \leq 0.6$. As another example it can be shown using eqs.(49) and
(50) that periodic orbits with $e<3\sqrt{5}/11=0.609836$ have $\varepsilon
<0.8$.

(iii) The Special Case of $e=1$

The maximum or boundary value for $e$ which is $e=1$ is also a special case
of interest. From its definition given by eq.(11), since $\kappa ^{2}\leq 1$%
, $e$ cannot be greater than $1$. In Appendix B, we show that $e=1$ always
gives an unbounded orbit for the periodic and the asymptotic periodic orbits
of Region I and the terminating orbits of Region II, but not the asymptotic
terminating orbit of Region I which has a radius given by eq.(47)
independent of $e$ and thus is not an unbounded orbit. Thus $e=1$ is the
boundary for $e$ for Region I (as well as for Region II). Many explicitly
simple relationships among $s,k,q_{\min },q_{1},$ etc. have been found on
the boundary line $e=1$, and they are given and proved in Appendix B. In
particular, we have, on $e=1$ in Region I, that

\begin{equation}
s^{2}=\frac{k^{2}}{4(1+k^{2})^{2}},
\end{equation}

\begin{equation}
\gamma =\left( \frac{1}{4(1+k^{2})}\right) ^{1/2},
\end{equation}

\begin{equation}
q_{\min }=\frac{1+k^{2}}{k^{2}},
\end{equation}

and

\begin{equation}
q_{1}=1+k^{2}.
\end{equation}

Examples of unbounded orbits for $e=1$ are shown in Fig.4 (a3-g3).

We shall now describe Regions II and II' for the orbit equation (36) given
by solution B for the case $\Delta >0$.

\section{Region II}

Consider the orbits expressed by eq.(36) given by solution B and
characterized mathematically by $\Delta >0$. The associated values for $%
(e,s) $ in this case satisfy $s>s_{1}$, where $s_{1}$ is the upper boundary
of Region I given by eq.(40). This region of parameter space defined by $%
s>s_{1} $ can be naturally divided into two sectors which we call Region II
and II' with Region II bordering Region I (see Fig.2). The boundary between
Regions II and II' is determined by the Schwarzschild radius in a manner to
be described later in this section.

We first want to prove that the lower boundary (for $s$) of Region II,
characterized by $\Delta =0$ as it is for the upper boundary of Region I,
also gives $k^{2}=1$, where $k^{2}$ is calculated from eq.(35) for Solution
B (In Section 3, we showed that for $k^{2}$ calculated from eq.(22) for
Solution A, $k^{2}=1$ implies $\Delta =0$). Substituting $\Delta =0$ into
eq.(32) gives

\begin{equation*}
A=B=\frac{1}{2}\sqrt[3]{g_{3}}=-\frac{1}{2}\sqrt{\frac{g_{2}}{3}}.
\end{equation*}

After noting that $A(=B)$ is a negative value for the range of $s$ values
for $\Delta =0$, substituting the above into eq.(35) gives $k^{2}=1$. We
also find from eq.(34) that $\gamma ^{2}=-3\sqrt[3]{g_{3}}/2=\sqrt{3g_{2}}/2$
which agrees with the $\gamma $ given by eq.(44), and we find from eq.(33)
that

\begin{equation}
a=\sqrt[3]{g_{3}}=-\sqrt{\frac{g_{2}}{3}}.
\end{equation}

Substituting these into eq.(36) gives the same orbit equation (43) for the
terminating orbit in Region II on its lower boundary as that for the
asymptotic periodic orbit in Region I on its upper boundary. Thus on the
boundary $k^{2}=1$ the equation for the orbits in Region II does not
represent a terminating orbit but is the same as the asymptotic periodic
orbit for Region I given by eq.(43) (see Fig.4, g1-g3). Also, from eqs.(42)
and (55), we see that the smallest root in eq.(23) in Solution (A) is
identified with the real root given by eq.(33) of Solution (B), i.e. $%
e_{3}=a $. Thus from eqs.(26) and (37), $q_{2}=q_{\max }$ when $k^{2}=1$,
i.e. the initial distance $q_{2}$ of the terminating orbit in Region II can
be identified as the continuation of $q_{\max }$ of the periodic orbit from
Region I. On the boundary of Regions I and II, the two other real roots $%
e_{1}=e_{2}$ given by eq.(42) of the cubic equation (16) agree with $b=$ $%
\overline{b}$ given below eq.(33). The line $k^{2}=1$ defined by eq.(40) is
the boundary between Regions I and II; it is the upper boundary for Region I
and is the lower boundary for Region II (see Fig.2). The above discussion
also illustrates the transition that takes place: from a periodic orbit to
an asymptotic periodic orbit to a terminating orbit, as one crosses the
boundary from Region I to II.

We now consider the upper boundary of Region II. We define this boundary to
be that obtained by requiring the planet's initial position to be just at
the Schwarzschild horizon, i.e. that obtained by setting $q=1$ initially at $%
\phi =0$. Setting $q=1$ in eq.(36) for $\phi =0$ which is $1/q=1/3+4a$, we
require $a=1/6$, where $a$ is the real root of the cubic equation (16). We
then use the equation 
\begin{equation}
4\left( \frac{1}{6}\right) ^{3}-\left( \frac{1}{6}\right) g_{2}-g_{3}=0,
\end{equation}

and substitute the expressions for $g_{2}$ and $g_{3}$ given in eq.(10) into
eq.(56) and solve for $s$. We find

\begin{equation}
s_{2}^{2}=\frac{1}{1-e^{2}}
\end{equation}

which we shall use as the equation for the upper boundary of Region II, for $%
0\leq e\leq 1$. Thus Region II is a region bounded between $e=0$ and $e=1$,
and between $s_{1}$ given by eq.(40) (the lower heavy solid line in Fig.2)
and $s_{2}$ given by eq.(57) (the upper heavy solid line in Fig.2), i.e. $%
s_{1}<s\leq s_{2}$. The region defined by $s_{2}<s\leq \infty $ and bounded
between $e=0$ and $e=1$ will be called Region II', for which the planet's
initial position ranges from just inside the Schwarzschild horizon up to the
center of the blackhole. Since the same terminating orbit equation (36)
applies in Regions II and II', the division into two regions may seem
unnecessary. However, the Schwarzschild radius is of physical significance,
and it is useful to know the location of the curve $s_{2}$ in the $(e,s)$
plot which indicates that the initial position of the planet is at the
Schwarzschild horizon. Separating out Region II' also makes it possible to
realize and appreciate that a very large region of the characterizing
parameter $s_{2}<s\leq \infty $ is of relevance only to a very small
physical region $0\leq q<1$ for the case where the initial position of the
planet is inside the Schwarzschild horizon.

For Region II, as $s$ increases its value above those on its lower boundary $%
s=s_{1}$ on which $k^{2}=1$, the value of $k^{2}$ calculated from eq.(35)
decreases from $1$. The curves of constant $k^{2}$ for $k^{2}=0.9,0.8,...$
can be easily obtained from eq.(35) where $A$ and $B$ are expressed in terms
of $s$ and $e$ (again using MAPLE FSOLVE) and they are presented in Fig.2.
However, the value of $k^{2}$ has a minimum value that is not $0$ in Region
II. First, we show in Appendix C that the $k^{2}=1/2$ curve is given by

\begin{equation}
s^{2}=\frac{1}{6(1-e^{2})}\left( 1+\sqrt{\frac{1+2e^{2}}{3}}\right) .
\end{equation}

The significance of this $k^{2}=1/2$ curve is that on it, as $e\rightarrow 1$%
, $s\rightarrow \infty $, just like the curve for the upper boundary of
Region II represented by eq.(57). For $k^{2}>0.5$, the constant $k^{2}$
curves intersect the $e=1$ line at some finite value of $s$, whereas for $%
k^{2}<0.5$, the constant $k^{2}$ curves intersect the upper boundary curve
given by eq.(57) at points for which the values of $e$ are less than $1$. We
then find that the minimum value of $k^{2}$ in Region II is equal to $%
1/2-1/(2\sqrt{5})=0.276393$ which is obtained by setting $e=0,s=1$ in
eqs.(32) and (35), and this value of $k^{2}$ appears at one coordinate point
only at $e=0$ and $s=1$. There is no orbit whose $k^{2}$ is less than $%
0.276393$ in Region II (see Fig.2), and $k^{2}$ is thus restricted to the
range $0.276393\leq k^{2}\leq 1$.

In Table 9, we present the coordinates $(e,s)$ of these curves of constant $%
k^{2}$ between $0.276393$ and $1$. In Table 10, we present the values of $%
q_{2}$ given by eq.(37), the initial distance of the planet from the
blackhole. Note that unlike $q_{1}$ for the terminating orbits in Region I
whose range is finite and small, $q_{2}$ can be infinite (for $e=1$ and $%
k^{2}>0.5$). Like the terminating orbits of Region I, the terminating orbits
of Region II can be characterized by $q_{2}$ and the angle $\phi _{2}$ at
which the planet enters the center of the blackhole. If we define the
"precession angle" $\Delta \phi $ for the terminating orbits as in eq.(25),
with $k$ and $\gamma $ defined by eqs.(35) and (34), then $\phi
_{2}=K(k)/\gamma =\Delta \phi /2+\pi $, or

\begin{equation}
\frac{\phi _{2}}{\pi }=\frac{1}{2}\frac{\Delta \phi }{\pi }+1.
\end{equation}

In Table 11, we present the values of $\phi _{2}$. Tables 10 and 11 are to
be used in conjunction with Table 9 that give the coordinates of the
constant $k^{2}$ curves. Examples of these terminating orbits obtained from
eq.(36) are shown in Fig.7(b)-(d). Again the dotted line shows the
continuation of the orbit beyond $\phi _{2}$. A planet coming from very far
away, i.e. an unbounded orbit with $e=1$, with an initial trajectory
perpendicular to the line joining it to the blackhole, can terminate at the
blackhole; the condition for this to happen is $s>0.25$. Figures 8 (a)-(c)
show three unbounded orbits $(e=1)$ as $s$ increases from just below to just
above the critical field parameter $s=0.25$. Figure 8(a) also shows an
example of a precession angle in which the planet makes more than three
revolutions around a blackhole before assuming a distance equal to its
initial distance (which is infinity) from the blackhole. As noted after
eq.(25), the actual precession angle in this case should be more
appropriately given by $2K(k)/\gamma -6\pi $ which can be obtained from the
presented value of $\Delta \phi /\pi =4.6378$ (where $\Delta \phi $ is
defined by eq.(25)) and gives $0.6378\pi $. Thus for Fig.8(a), $0.6378\pi $
gives the angle between the initial incoming trajectory from very far away
at $\phi =0$ and the final outgoing trajectory going to infinity, i.e. $%
0.6378\pi $ is the polar angle of the direction of the outgoing trajectory
going to infinity with respect to the $x$-axis (but we have not extended the
outgoing trajectory far enough to show the accuracy of this angle).

We now present some useful simple expressions for the following special
cases.

(i) Special Case on the Upper Boundary of Region II Given By Eq.(57)

We show in Appendix D that on the upper boundary of Region II given by
eq.(57), the values of $k^{2}$ and $\gamma $ given by eqs.(35) and (34)
become

\begin{equation}
k^{2}=\frac{1}{2}-\frac{1}{4\sqrt{1/3-g_{2}}},
\end{equation}

\begin{equation}
\gamma =\left[ \frac{1}{4}\left( \frac{1}{3}-g_{2}\right) \right] ^{1/4},
\end{equation}

where the $s$ values for $g_{2}$ are given by eq.(57).

(ii) Special Case on the Right Boundary $e=1$ of Region II

Just as for Region I, there are simple and interesting relations among $%
k^{2} $, $s$ and $\gamma $ on the right boundary $e=1$ of Region II, and
they are shown in Appendix D. In particular, we have, on $e=1$ in Region II,
that for $s>1/4$,

\begin{equation}
k^{2}=\frac{1}{2}+\frac{1}{8s},
\end{equation}

or that, for $1\geq k^{2}>1/2$,

\begin{equation}
s=\frac{1}{8(k^{2}-1/2)},
\end{equation}

and that

\begin{equation}
\gamma =\sqrt{\frac{s}{2}}.
\end{equation}

Appendix D also presents a special case given by $s^{2}=1/12$ that is
notable.

\section{Region II'}

While we may call the entire sector $s>s_{1}$ given by eq.(40) above Region
I in Fig.2 just one region that allows only terminating orbits given by
eq.(36), it is useful to divide it into Regions II and II' using the curve $%
s=s_{2}$ given by eq.(57). Region II' is the region of parameter space in $%
(e,s)$ for which $s>s_{2}$ and $0\leq e<1$. The heavy solid curve labeled $%
s_{2}$ in Fig.2 delineates the boundary of Region II' which separates it
from Region II. Despite the apparent large size of Region II', the
terminating orbits here have little variety in the sense that the range of
initial distances $q_{2^{\prime }}$ that are given by $1/q_{2^{\prime
}}=1/3+4a$ (see eq.(37)), is limited $(0\leq q_{2^{\prime }}<1)$ and the
range of the angle $\phi _{2^{\prime }}=K(k)/\gamma $ at which the planet
enters the blackhole is also limited. It can be shown that the range of $%
\phi _{2^{\prime }}$ is $0\leq \phi _{2^{\prime }}<0.789\pi $. An example of
a terminating orbit obtained from eq.(36) in Region II' is shown as the
solid line in Fig.7(e); the dotted line shows the continuation of the orbit
beyond $\phi _{2^{\prime }}$. It may be of some mathematical interest to
note that as $s\rightarrow \infty $ in Region II', the modulus of the
Jacobian elliptic functions used to describe the orbits does not go to zero;
instead $k^{2}\rightarrow (2-\sqrt{3})/4=0.0669873$, and thus $k^{2}$ in
Region II' is restricted to the range $0.0669873\leq k^{2}<0.5$.

Tables 1-8, the orbit equations (20), (29) and (36), and the description of
the orbits and the three regions where these orbit equations apply, complete
our characterization of all possible planetary orbits in the Schwarzschild
geometry.

We now briefly discuss how all this may be used for the Kerr geometry when
the spinning blackhole has a spin angular momentum per unit mass of the
blackhole that is relatively small compared to the orbital angular momentum
per unit mass of the planet.

\section{Kerr Geometry}

The spinning blackhole is assumed to have a spin angular momentum $J$ given
by [1]

\begin{equation}
J=Mac,
\end{equation}

where $ac$ can be identified as the spin angular momentum per unit mass of
the blackhole and is the quantity to be compared with $h$, the orbital
angular momentum per unit mass of the planet. The Kerr geometry becomes the
Schwarzschild geometry in the limit $ac/h\rightarrow 0$.

The worldline of a particle moving in the equatorial plane $\theta =\pi /2$
satisfies the equations [1]

\begin{equation}
\overset{\cdot }{t}=\frac{1}{D}\left[ \left( r^{2}+a^{2}+\frac{\alpha a^{2}}{%
r}\right) \kappa -\frac{\alpha ah}{cr}\right] ,
\end{equation}

\begin{equation}
\overset{\cdot }{\phi }=\frac{1}{D}\left[ \frac{\alpha ac\kappa }{r}+\left(
1-\frac{\alpha }{r}\right) h\right] ,
\end{equation}

where $D\equiv r^{2}-\alpha r+a^{2}$. For the equatorial trajectories of the
planet in the Kerr geometry, the combined energy equation is

\begin{equation}
\overset{\cdot }{r}^{2}+\frac{h^{2}-a^{2}c^{2}(\kappa ^{2}-1)}{r^{2}}-\frac{%
\alpha (h-ac\kappa )^{2}}{r^{3}}-\frac{c^{2}\alpha }{r}=c^{2}(\kappa ^{2}-1).
\end{equation}

Provided that $a^{2}/\alpha ^{2}<1$ and $ac\kappa /h<<1$, to the first order
in $ac\kappa /h$, it is not difficult to see, by comparing eqs.(66)-(68)
with eqs.(3)-(6), that we can re-scale $\alpha $ to $\alpha ^{\prime
}=\alpha (1-2ac\kappa /h)$, $s$ to $s^{\prime }=s(1-ac\kappa /h)$, and $\phi 
$ to $\phi ^{\prime }=\phi \lbrack 1-b(ac\kappa /h)]$, where $b$ is some
approximation constant, such that the results we have presented for the
orbits in the Schwarzschild geometry are approximately applicable for the
orbits in the Kerr geometry in terms of the scaled parameters. That is, the
orbits in the equatorial plane and their characterization for the
Schwarzschild and Kerr geometries are qualitatively very similar to the
first order in $ac\kappa /h$ except that the basic parameters $s$, $\alpha $
and $\phi $ have to be slightly rescaled. Levin and Perez-Giz [5] obtained
their orbits in the Kerr geometry from numerically integrating eqs.(66)-(68)
and it would be interesting to study and examine when and how the planet's
orbits in the Kerr geometry that they obtained can be related with our
results with the scaled parameters, and when and how they begin to differ
significantly from those in the Schwarzschild geometry that we presented in
this paper.

\section{Trajectory of Light}

We now consider the deflection of light by a gravitational field. We cannot
use the proper time $\tau $ as a parameter. So we use some affine parameter $%
\sigma $ along the geodesic [1]. Considering motion in the equatorial plane,
the geodesic equations give eqs.(3) and (5), and we replace the $r$-equation
(4) by the first integral of the null geodesic equation, and we have [1]

\begin{equation}
\left( 1-\frac{\alpha }{r}\right) \overset{\cdot }{t}=\kappa ,
\end{equation}

\begin{equation}
c^{2}\left( 1-\frac{\alpha }{r}\right) \overset{\cdot }{t}^{2}-\left( 1-%
\frac{\alpha }{r}\right) ^{-1}\overset{\cdot }{r}^{2}-r^{2}\overset{\cdot }{%
\phi }^{2}=0,
\end{equation}

\begin{equation}
r^{2}\overset{\cdot }{\phi }=h,
\end{equation}

where the derivative $\overset{\cdot }{}$ represents $d/d\sigma $.
Substituting eqs.(69) and (71) into (70) gives the 'combined' energy equation

\begin{equation}
\overset{\cdot }{r}^{2}+\frac{h^{2}}{r^{2}}\left( 1-\frac{\alpha }{r}\right)
=c^{2}\kappa ^{2}.
\end{equation}

Substituting $dr/d\sigma =(dr/d\phi )(d\phi /d\sigma )=(h/r^{2})(dr/d\phi )$
and $u=1/r$ into the combined energy equation gives the differential
equation for the trajectories of light in the presence of a gravitational
field

\begin{equation}
\left( \frac{du}{d\phi }\right) ^{2}=\alpha u^{3}-u^{2}+\frac{c^{2}\kappa
^{2}}{h^{2}}.
\end{equation}

The constants $\kappa $ and $h$ have a physical significance through their
ratio $\kappa /h$ as follows: Let $R$ denote the distance of the light beam
to the center of a star or blackhole when the trajectory of the light beam
is such that $du/d\phi =0$. $R$ can either be associated with the distance
of closest approach of the light beam to the blackhole or with the initial
distance to the blackhole of the light beam. The latter case is associated
with light trajectories that terminate at the blackhole. With $R$ so defined
and letting $u_{1}\equiv 1/R$, we can set $c^{2}\kappa ^{2}/h^{2}$ to be
equal to $u_{1}^{2}-\alpha u_{1}^{3}$ [8].

It is again convenient to consider the problem in terms of the dimensionless
inverse distance $U$ defined by

\begin{equation}
U=\frac{\alpha }{r}=\alpha u=\frac{1}{q}.
\end{equation}

$U$ defined here is slightly different from the $U$ defined by eqs.(8) and
(19) previously. In terms of $U$ of eq.(74), eq.(73) becomes

\begin{equation}
\left( \frac{dU}{d\phi }\right) ^{2}=U^{3}-U^{2}+\frac{c^{2}\kappa
^{2}\alpha ^{2}}{h^{2}}.
\end{equation}

Since $dU/d\phi =0$ at $r=R$, one root $U$ which we call

\begin{equation}
U_{1}\equiv \frac{\alpha }{R}\equiv \alpha u_{1}
\end{equation}

of the cubic equation $U^{3}-U^{2}+c^{2}\kappa ^{2}\alpha ^{2}/h^{2}=0$ is
known, and the term $c^{2}\kappa ^{2}\alpha ^{2}/h^{2}$ on the right hand
side of eq.(75) can be replaced by $-U_{1}^{3}+U_{1}^{2}$, and the other two
roots of the cubic equation $U^{3}-U^{2}-U_{1}^{3}+U_{1}^{2}=0$ can be found
from solving a quadratic equation. We denote the three roots of the cubic
equation by $e_{1},e_{2},e_{3}$. Thus writing eq.(75) as

\begin{equation}
\left( \frac{dU}{d\phi }\right) ^{2}=U^{3}-U^{2}-U_{1}^{3}+U_{1}^{2}
\end{equation}

the trajectory of light represented by an equation for $U$ as a function of
the polar angle $\phi $ obtained from integrating eq.(77) can be
characterized by a single parameter $U_{1}$ which essentially specifies
either the distance of the closest approach or the initial distance of the
light beam to the blackhole (These distances are scaled by the Schwarzschild
radius of the blackhole). As in our discussion of the planets, our
references to the initial position of the light beam assume that the
trajectory of the light beam at that initial position is perpendicular to
the line joining that position to the star or blackhole. The range of $U_{1}$
is clearly between $0$ and $\infty $, where $U_{1}=0$ means that the light
beam is infinitely far away from the blackhole, $U_{1}=1$ means that the
light beam is at the Schwarzschild radius at its closest approach or its
initial position, and $U_{1}=\infty $ means that the light beam is at the
center of the blackhole. As we show in the following, the region $0\leq
U_{1}\leq \infty $ can be appropriately divided into three sectors which we
again call Regions I, II and II'. The similarity between the
characterization of these three regions with that for the planetary orbits
discussed in the previous sections will become apparent. Not surprisingly
perhaps, only a single parameter which we choose to be $U_{1}$, is needed
for the characterization of the trajectories of a light beam in contrast to
the two parameters (which we choose to be $e$ and $s$), which we needed for
the characterization of the orbits of a planet. The relationship between $%
U_{1}$ and $R$, from eqs.(76) and (2), is

\begin{equation*}
R=\frac{2}{U_{1}}\left( \frac{GM}{c^{2}}\right) .
\end{equation*}

Region I: $0\leq U_{1}\leq 2/3,$ or $\infty >R\geq 3GM/c^{2}$

Here $R$ denotes the distance of closest approach of a light beam that comes
from a great distance. We let

\begin{eqnarray}
e_{1} &=&\frac{1}{2}[1-U_{1}+(1+2U_{1}-3U_{1}^{2})^{1/2}],  \notag \\
e_{2} &=&U_{1},  \notag \\
e_{3} &=&\frac{1}{2}[1-U_{1}-(1+2U_{1}-3U_{1}^{2})^{1/2}],
\end{eqnarray}

with $e_{1}>e_{2}>e_{3},$ and we consider the region $e_{1}>e_{2}>U\geq
e_{3} $, and write eq.(77) as

\begin{equation}
\left( \frac{dU}{d\phi }\right) ^{2}=(e_{1}-U)(e_{2}-U)(U-e_{3}).
\end{equation}

Equation (79) can be integrated [7] with $\phi $ expressed in terms of an
inverse $sn$ function. After a little algebra and re-arrangement, we find
the trajectory's equation in terms of the Jacobian elliptic functions of
modulus $k$ to be

\begin{equation}
\frac{1}{q}=\frac{(e_{1}-e_{3})e_{2}-(e_{2}-e_{3})e_{1}sn^{2}(\gamma \phi ,k)%
}{(e_{1}-e_{3})-(e_{2}-e_{3})sn^{2}(\gamma \phi ,k)},
\end{equation}

where

\begin{eqnarray}
\gamma &=&\frac{(e_{1}-e_{3})^{1/2}}{2},  \notag \\
k^{2} &=&\frac{e_{2}-e_{3}}{e_{1}-e_{3}}.
\end{eqnarray}

The angle of deflection $\Delta \phi $ can be obtained as follows. If we set 
$q=\infty $ and also set $\phi =\pi /2+\Delta \phi /2$ as the incoming angle
in eq.(80) (see Fig.9a for the special case of $\Delta \phi /2=45^{\circ }$)
where $\Delta \phi $ denotes the total angle of deflection of light by the
mass $M$, we get the following equation for determining $\Delta \phi $
exactly:

\begin{equation*}
sn^{2}\left[ \gamma (\frac{\pi }{2}+\frac{\Delta \phi }{2}),k\right] =\frac{%
(e_{1}-e_{3})e_{2}}{(e_{2}-e_{3})e_{1}},
\end{equation*}

where $e_{1},e_{2},e_{3},\gamma ,k,$ are given by eqs.(78) and (81). It can
also be expressed as

\begin{equation}
\Delta \phi =-\pi +\frac{2}{\gamma }sn^{-1}(\psi ,k),
\end{equation}

where

\begin{equation*}
\psi =\left[ \frac{(e_{1}-e_{3})e_{2}}{(e_{2}-e_{3})e_{1}}\right] ^{1/2}
\end{equation*}

Equations (80)-(82) were first given by one of us in ref.6. Examples of
these trajectories obtained from eq.(80) are presented in polar coordinates $%
(q,\phi )$ in Fig.9, where the blackhole is located at the origin. By
setting the angles of deflection $\Delta \phi $ presented in Fig.9 to be $%
\pi /2,\pi ,3\pi /2,2\pi $, the corresponding values of $U_{1}$ can be
determined from eqs.(82) and (78) using the MAPLE FSOLVE program, and they
are found to correspond to the distances of closest approach $%
R=4.6596GM/c^{2}$, $3.5206GM/c^{2}$, $3.2085GM/c^{2}$, $3.0902GM/c^{2}$
respectively. The case of $R=3.5206GM/c^{2}$ is interesting as it
corresponds to the light ray being turned around by $180^{\circ }$. That the
upper boundary of Region I characterized by $U_{1}=2/3$ or $R=3GM/c^{2}$ is
a very special case can be seen mathematically because it results in $%
e_{1}=e_{2}=2/3,e_{3}=-1/3,$ and hence $k^{2}=1$, $\gamma =1/2$ and $%
U=2/3=const.$ from eqs.(81) and (80). Physically, it results in the light
circling the blackhole with a radius $R=3GM/c^{2}$ even though the
trajectory has been shown to be an unstable one [1]. This known result can
also be simply obtained from the equation of motion $d^{2}U/d\phi
^{2}=(3/2)U^{2}-U$ for $U=const.$ and thus $U=2/3$. If one compares the size
of the unstable circular photon orbit with the allowed limiting radii of the
planetary asymptotic periodic orbits $(2\leq q_{\min }\leq 2.25$ or $%
4GM/c^{2}\leq r_{\min }\leq 4.5GM/c^{2})$, one can see that the radius of
the asymptotic circular path of a planet around a blackhole is still a
little larger than that for a photon, but not by much.

The lower boundary of Region I characterized by $U_{1}=0$ or $R=\infty $
gives $e_{1}=1,e_{2}=e_{3}=0$, $k^{2}=0$ and $\gamma =1/2$, and thus gives $%
U=0$ or $r=\infty $ which is a limiting case as the light ray that is
infinitely far away at its closest approach to the blackhole is completely
undeflected.

As in the case of the Region I particle orbits discussed in Section 3, the
squared modulus $k^{2}$ of the elliptic functions that describe the
trajectories of light here also covers the entire range $0\leq k^{2}\leq 1$;
it varies from $0$ at the lower boundary to $1$ at the upper boundary.

For small $U_{1}$, the trajectory of light given by eq.(80) has been shown
[6] to reduce to

\begin{equation}
\frac{1}{r}\simeq \frac{\cos \phi }{R}+\frac{GM}{c^{2}R^{2}}(1+\cos \phi
+\sin ^{2}\phi ),
\end{equation}

and the total deflection of light to reduce to the well known result

\begin{equation}
\Delta \phi \simeq \frac{4GM}{c^{2}R}.
\end{equation}

It can be shown from our exact result given by eq.(82) that this approximate
expression (84) still gives an accuracy of two significant figures for $%
U_{1}=0.1$ or $R=20GM/c^{2}$.

As $U_{1}$ approaches $2/3$, or as $R$ approaches $3GM/c^{2}$, we may let $%
U_{1}=2/3-\delta $, where $\delta \equiv (2/3)(1-3GM/c^{2}R)$ is a small
positive number. From eqs.(78) and (81), we can express the quantities $%
2/\gamma $, $\psi $ and $k$ appearing in eq.(82) in power series in $\delta $
and find, to the first order in $\delta $, $2/\gamma \simeq 4(1-\delta
/2+...)$, $\psi \simeq 1-\delta /2+...$, and $k\simeq 1-\delta +...$.
Substituting these into eq.(82) immediately gives an expression for $\Delta
\phi $ which is correct to the first order in $\delta $. If an attempt is
made to find an expansion of $sn^{-1}(\psi ,k)$ near $k=1$, since $%
sn^{-1}(\psi ,1)=\tanh ^{-1}\psi =\ln [(1+\psi )/(1-\psi )]^{1/2}$, the
expansion would involve terms in $\ln \delta $ (which is a large number for
small $\delta $) and ordering the expansion terms in the right way can be
tricky. Different forms of such expansions have been given and studied by
various authors [9]. As we showed above and in Fig.9, our exact expressions
given by eqs.(80) and (82) can be used simply and directly for all cases in
Region I.

As $U_{1}$ increases beyond $2/3$ or as the distance of closest approach $R$
of the light beam to the blackhole becomes smaller than $3GM/c^{2}$, the
light is not just deflected but is absorbed by and terminates at the
blackhole. It is useful to divide the region $2/3<U_{1}\leq \infty $ or $%
3GM/c^{2}>R\geq 0$ into two regions that we call Region II and II' that are
separated by the Schwarzschild horizon, as we discuss below. Region II is
for $R$ from $3GM/c^{2}$ up to the Schwarzschild horizon, and Region II' is
for $R$ from the Schwarzschild horizon up to the center of the blackhole.

Region II: $2/3<U_{1}\leq 1,$ or $3GM/c^{2}>R\geq 2GM/c^{2}$

Here $R$ denotes the initial distance to the blackhole of the light beam
which has initial trajectory (as $\phi $ increases from $0$) perpendicular
to the line joining it to the blackhole.

As $U_{1}$ increases beyond $2/3$, $U_{1}$ becomes greater than $%
[1-U_{1}+(1+2U_{1}-3U_{1}^{2})^{1/2}]/2$, and the order of the three roots
must be changed to maintain the inequality $e_{1}>e_{2}>e_{3}$. We write

\begin{eqnarray}
e_{1} &=&U_{1},  \notag \\
e_{2} &=&\frac{1}{2}[1-U_{1}+(1+2U_{1}-3U_{1}^{2})^{1/2}],  \notag \\
e_{3} &=&\frac{1}{2}[1-U_{1}-(1+2U_{1}-3U_{1}^{2})^{1/2}].
\end{eqnarray}

We consider the region $U>e_{1}>e_{2}>e_{3}$, and write eq.(77) as

\begin{equation}
\left( \frac{dU}{d\phi }\right) ^{2}=(U-e_{1})(U-e_{2})(U-e_{3}).
\end{equation}

Equation (86) can be integrated [7] with $\phi $ expressed in terms of an
inverse $sn$ function. After some rearrangement, we find

\begin{equation}
\frac{1}{q}=\frac{e_{1}-e_{2}sn^{2}(\gamma \phi ,k)}{cn^{2}(\gamma \phi ,k)},
\end{equation}

where $\gamma $ and $k^{2}$ are calculated using the same expressions given
by eq.(81) but with $e_{1},e_{2},e_{3}$ given by eq.(85).

The expressions for $e_{1},e_{2},e_{3}$ given by eqs.(78) and (85) coincide
at $k^{2}=1$ for which $e_{1}=e_{2}=2/3,e_{3}=-1/3$, and both equations (80)
and (87) give $U=2/3$ or $r=3GM/c^{2}$ independent of $\phi $.

Eq.(87) gives a trajectory of light which terminates at the blackhole when $%
\phi =\phi _{2}=K(k)/\gamma $. As in our discussion of the terminating
orbits for the planet, the terminating light ray trajectories can be
characterized by the angle $\phi _{2}$ with which the light beam enters the
center of the blackhole.

As $U_{1}$ increases from $2/3$ to $1$, $k^{2}$ covers the entire range $%
1\geq k^{2}\geq 0$; it decreases from $1$ to $0$. When $U_{1}=1$, i.e. when
the light beam grazes the Schwarzschild horizon, $%
e_{1}=1,e_{2}=e_{3}=0,k^{2}=0,\gamma =1/2$, and we have the trajectory of
light given by

\begin{equation}
\frac{1}{q}=\frac{1}{\cos ^{2}(\phi /2)},
\end{equation}

which gives, for $\phi =0,U=1$ or $r=\alpha $, and for $\phi =\pi ,U=\infty $
or $r=0$, i.e. the light is absorbed at the center of the blackhole.
Examples of the trajectories of light obtained from eqs.(87) and (88) for $%
U_{1}=5/6=0.83333$ ($R=2.4GM/c^{2}$) and $1$ ($R=2GM/c^{2}$) in Region II
are shown as the solid lines in Fig.10 (a) and (b). The path that emerges
from the center of the blackhole when $\phi $ is continued beyond $\phi _{2}$
(shown as a dotted line in Fig.10) again may be interesting if the concept
of whitehole is of any physical relevance.

When the distance $R$ to the blackhole at $\phi =0$ is inside the
Schwarzschild horizon, the terminating path takes on a somewhat different
form as we show below.

Region II': $1<U_{1}\leq \infty ,$ or $2GM/c^{2}>R\geq 0$

Here $R$ has the same meaning as that in Region II. As $U_{1}$ increases
beyond $1$, i.e. when $R$ is less than the Schwarzschild radius, $e_{1}$ in
eq.(85) remains real while $e_{2}$ and $e_{3}$ become complex. We now write
the three roots of the cubic equation $U^{3}-U^{2}-U_{1}^{3}+U_{1}^{2}=0$ as 
$a$, $b$ and $\overline{b}$ given by

\begin{eqnarray}
a &=&U_{1},  \notag \\
b &=&\frac{1}{2}[1-U_{1}+i(3U_{1}^{2}-2U_{1}-1)^{1/2}],  \notag \\
\overline{b} &=&\frac{1}{2}[1-U_{1}-i(3U_{1}^{2}-2U_{1}-1)^{1/2}].
\end{eqnarray}

We consider the region $U>a$, and write eq.(77) as

\begin{equation}
\left( \frac{dU}{d\phi }\right) ^{2}=(U-a)(U-b)(U-\overline{b}).
\end{equation}

This equation can be integrated [7] with $\phi $ expressed in terms of an
inverse $cn$ function. After a little algebra, we find

\begin{equation}
\frac{1}{q}=\frac{\gamma ^{2}+a-(\gamma ^{2}-a)cn(\gamma \phi ,k)}{%
1+cn(\gamma \phi ,k)},
\end{equation}

where

\begin{equation}
\gamma =[U_{1}(3U_{1}-2)]^{1/4}
\end{equation}

and

\begin{equation}
k^{2}=\frac{1}{2}-\frac{3U_{1}-1}{4\sqrt{U_{1}(3U_{1}-2)}}=\frac{1}{2}-\frac{%
3a-1}{4\gamma ^{2}}.
\end{equation}

Equation (91) gives the trajectory of light when $R$ is inside the
Schwarzschild horizon and it terminates at the blackhole when $\phi =\phi
_{2^{\prime }}=2K(k)/\gamma $, where $k$ and $\gamma $ are given by eqs.(93)
and (92). On the boundary with Region II where $U_{1}=1$, and $k^{2}=0$, $%
\gamma =1$ from eqs.(93) and (92), eq.(91) becomes eq.(88) and thus there is
no discontinuity in the orbit as it makes a transition from Region II to
Region II' across $U_{1}=1$.

We note that as in the case of Region II' for the planetary orbits, Region
II' for light trajectories covers a semi-infinite range of the parameter
characterizing it ($1<U_{1}\leq \infty $) but is of relevance only to a very
small physical region $2GM/c^{2}>R\geq 0$ for the initial position of a
light beam inside the Schwarzschild horizon. The terminating orbits of light
rays in Region II' are also of very little variety as $\phi _{2^{\prime }}$
is restricted to a limited range of $0\leq \phi _{2^{\prime }}\leq \pi $. An
example of a terminating trajectory obtained from eq.(91) is shown as the
solid line in Fig.10(c) for $U_{1}=10$ ($R=0.2GM/c^{2}$); the dotted line
again represents a trajectory of light coming out from the center of the
blackhole as $\phi $ is continued beyond $\phi _{2^{\prime }}$. It may be of
some mathematical interest to note that as $U_{1}\rightarrow \infty $ or $%
R\rightarrow 0$, the squared modulus of the Jacobian elliptic functions used
to describe the trajectories $k^{2}$ approaches a value $(2-\sqrt{3}%
)/4=0.0669873$ \ that is the same as that given in Section 5 for the case of
Region II' for the planetary orbits. Thus the squared modulus of the
elliptic functions that describe the terminating light trajectories in
Region II' is restricted to a very small range $0<k^{2}\leq 0.0669873$ even
as Region II' consists of a very large interval $1<U_{1}\leq \infty $.

\section{Summary}

We have presented exact analytic expressions given by eqs.(20), (29) and
(36) for the planetary orbits in the Schwarzschild geometry. The equations
relate the distance $r$ of the planet from the star or blackhole to the
polar angle $\phi $ and are described explicitly by Jacobian elliptic
functions of modulus $k$. Equation (20) gives periodic orbits that describe
a planet precessing around a star or blackhole, while eqs.(29) and (36) give
terminating orbits that describe a planet plunging into the center of a
blackhole. One of the most important aspects of our analysis is the
construction of a map with coordinates $(e,s)$ that we use to view all
possible orbits in their entirety, where the two dimensionless parameters $e$
and $s$ are defined by eqs.(11) and (12) which we call the energy and field
parameters respectively. For $0\leq e\leq 1$, we show that there are three
regions which we call Regions I $(0\leq s\leq s_{1})$, II $(s_{1}<s\leq
s_{2})$ and II' $(s_{2}<s\leq \infty )$ where these orbits are applicable,
and where $s_{1}$ and $s_{2}$ that depend on $e$ are given by eqs.(40) and
(57) respectively (Fig.2). Region I has periodic and terminating orbits
given by eqs.(20) and (29). Regions II and II' have terminating orbits only
given by eq.(36). We have divided Region I into grids that consist of lines
of constant precession angle $0\leq \Delta \phi \leq \infty $ given by
eq.(25) and lines of constant true eccentricity $0\leq \varepsilon \leq 1$
defined by eq.(28) (Fig.3); the lines of constant $\Delta \phi $ are
obtained from solving eqs.(38) and (99), and those of constant $\varepsilon $
from solving eqs.(38) and (100). These grids make the identification of all
possible periodic orbits, including the unbounded and asymptotic ones,
convenient and precise. Numerous numerical results for orbits in Region I
are presented in Tables 1-6, and examples of precessing orbits, including
the unbounded ones, are shown in Figs.4 and 5. Among the interesting
results, for example, Table 6 for $\Delta \phi =2\pi ,$ $e=1,$ $s=0.248804$
and Fig.4 f3 show that a planet coming from infinity at zero polar angle,
makes a complete loop about the blackhole, and returns to infinity with a
polar angle that approaches $2\pi $. Figure 3, or a more refined version of
it that can be constructed using the expressions for $\Delta \phi $ and $%
\varepsilon $ that we presented, can be fruitfully used with the
experimental observation data. As for the terminating orbits, our orbit
equations assume the planet's initial trajectory to be perpendicular to the
line joining it to the star or blackhole. The terminating orbits of Region I
require the planet to be at a distance $q_{1}<2$ from the star or blackhole,
and those of Region II' requires $q_{2^{\prime }}\leq 1$, i.e. they require
the planet to be initially very close to the blackhole or within the
Schwarzschild horizon. The terminating orbits of Region II, on the other
hand, are more interesting as the planet can be initially at a distance $%
1<q_{2}\leq \infty $. We have shown how the periodic orbits of Region I
become the asymptotic periodic orbits as $s\rightarrow s_{1}$, and then
become terminating orbits as $s$ becomes greater than $s_{1}$. It is
interesting to note that a planet initially at infinity with $e=1$ and in a
trajectory perpendicular to the line joining it to a blackhole can be
absorbed by the blackhole if $s$ is greater than $0.25$.

We have also presented exact analytic expressions given by eqs.(80), (87)
and (91) for the trajectories of light in the presence of a star or
blackhole depending on the value of one parameter $U_{1}$ that has a range
which can be divided into three regions: Regions I $(0\leq U_{1}\leq 2/3)$,
II $(2/3<U_{1}\leq 1)$ and II' $(1<U_{1}\leq \infty )$, where $U_{1}$ is
defined by eq.(76). In Region I, the deflection of light can range from
small angles to going continuously around the star or blackhole in a circle.
In Regions II and II', light is absorbed into the center of the blackhole.
Among the interesting results, a deflection of light by $180^{\circ }$
requires a distance of closest approach $R$ to the blackhole equal to $%
3.5206GM/c^{2}$ (Fig.9b), and for $R<3GM/c^{2}$ , light will be absorbed by
the blackhole.

We have thus presented a complete map that can help to identify
characteristics of stars and blackholes (that are not spinning too fast)
from the observed characteristics of objects or light beams that are
affected by them.

\section{References}

*Electronic address: fhioe@sjfc.edu

[1] M.P. Hobson, G. Efstathiou and A.N. Lasenby: General Relativity,
Cambridge University Press, 2006, Chapters 9 and 10.

[2] E.T. Whittaker: A Treatise on the Analytical Dynamics of Particles and
Rigid Bodies, 4th Edition, Dover, New York, 1944, Chapter XV.

[3] Y. Hagihara, Japanese Journal of Astronomy and Geophysics VIII, 67
(1931).

[4] S. Chandrasekhar: The Mathematical Theory of Black Holes, Oxford
University Press, 1983, Chapter 3.

[5] J. Levin and G. Perez-Giz, Phys. Rev. D 77, 103005 (2008).

[6] F.T. Hioe, Phys. Lett. A 373, 1506 (2009).

[7] P.F. Byrd and M.D. Friedman: Handbook of Elliptic Integrals for
Engineers and Scientists, 2nd Edition, Springer-Verlag, New York, 1971. See
in particular p.72, 74, 81, 86 for the formulas used in this paper.

[8] See e.g. J. L. Martin: General Relativity: A Guide to its Consequences
for Gravity and Cosmology, Ellis Horwood Ltd., Chichester,1988, Chapter 4.

[9] S.V. Iyer and A.O. Petters, Gen. Relat. and Grav. 39, 1563 (2007), C.R.
Keeton and A.O. Petters, Phys. Rev. D 72, 104006 (2005).

\section{Appendix A Relation Among $s,e$ and $k^{2}$}

In this Appendix, we derive the relation among $s,e$ and $k^{2}$ given by
eqs.(38) and (39). Substituting eq.(23) into eq.(22) and after a little
algebra, we find

\begin{equation}
\tan \frac{\theta }{3}=\frac{\sqrt{3}k^{2}}{2-k^{2}},
\end{equation}

and hence we find

\begin{eqnarray}
\sin \frac{\theta }{3} &=&\frac{\sqrt{3}k^{2}}{2\sqrt{1-k^{2}+k^{4}}}, \\
\cos \frac{\theta }{3} &=&\frac{2-k^{2}}{2\sqrt{1-k^{2}+k^{4}}}.
\end{eqnarray}

We then find

\begin{eqnarray}
\cos \left( \frac{\theta }{3}+\frac{4\pi }{3}\right) &=&\frac{-1+2k^{2}}{2%
\sqrt{1-k^{2}+k^{4}}},  \notag \\
\cos \left( \frac{\theta }{3}+\frac{2\pi }{3}\right) &=&\frac{-1-k^{2}}{2%
\sqrt{1-k^{2}+k^{4}}}.
\end{eqnarray}

Equations (96)-(98) and (23) allow $e_{1},e_{2},e_{3}$ to be expressed in
terms of $g_{2}$ and $k^{2}$ as%
\begin{eqnarray}
e_{1} &=&\sqrt{\frac{g_{2}}{12}}\frac{2-k^{2}}{\sqrt{1-k^{2}+k^{4}}},  \notag
\\
e_{2} &=&\sqrt{\frac{g_{2}}{12}}\frac{-1+2k^{2}}{\sqrt{1-k^{2}+k^{4}}}, 
\notag \\
e_{3} &=&\sqrt{\frac{g_{2}}{12}}\frac{-1-k^{2}}{\sqrt{1-k^{2}+k^{4}}}.
\end{eqnarray}

Substituting eq.(96) into the relation $\cos \theta =-3\cos (\theta
/3)+4\cos ^{3}(\theta /3)$, and substituting the result into eq.(24) gives
the relation among $s,e$ and $k^{2}$ given by eqs.(38) and (39). For $%
k^{2}=1 $, we get $\cos \theta =-1$ from eqs.(38) and (39), and from eq.(98)
we get $e_{1}=e_{2}=\sqrt{g_{2}/12}=-(1/2)e_{3}$.

From eqs.(25), (21), (98) and (10), we find an expression for the precession
angle $\Delta \phi $ given by eq.(25) in terms of $k$ and $s$ to be

\begin{equation}
\Delta \phi =4K(k)\left( \frac{1-k^{2}+k^{4}}{1-12s^{2}}\right) ^{1/4}-2\pi .
\end{equation}

\section{Appendix B The Energy Parameter $e$ and the True Eccentricity $%
\protect\varepsilon $ in Region I}

In this Appendix, we show the relation between the energy parameter $e$ and
the true eccentricity $\varepsilon $. The energy parameter $e$ in all three
regions I, II and II' is defined by eq.(11). The general or true
eccentricity $\varepsilon $ is defined by eq.(28). From eq.(28), (26) and
(27) and using the expressions (98) of $e_{1},e_{2},e_{3}$ in terms of $%
k^{2} $ given in Appendix A, we find that $\varepsilon $ can be expressed as

\begin{equation*}
\varepsilon =\frac{6(e_{2}-e_{3})}{1+6(e_{2}+e_{3})}
\end{equation*}

or

\begin{equation}
\varepsilon =\frac{3k^{2}\sqrt{1-12s^{2}}}{2\sqrt{1-k^{2}+k^{4}}-(2-k^{2})%
\sqrt{1-12s^{2}}}.
\end{equation}

For small $s^{2}$ and $k^{2}$, we find that

\begin{equation}
\varepsilon \simeq \frac{3k^{2}/2}{1-(1-6s^{2})}\simeq \frac{k^{2}}{4s^{2}}%
\simeq e
\end{equation}

because $k^{2}\simeq 4es^{2}$ for very small $k^{2}$ and $s^{2}$ [6]. We
thus confirm the identification of $e$ defined in eq.(11) with the
eccentricity of the orbit in Newtonian mechanics. As we pointed out in the
text, $\varepsilon $ is generally not equal to $e$. However, as we show
below, $\varepsilon =e$ exactly when $e=1$ and in this case the orbits are
unbounded.

For the possibility of unbounded orbits in Region I, we set the initial $%
q=\infty $ at $\phi =0$ in eq.(20), and get

\begin{equation}
\frac{1}{3}+4e_{3}=0,
\end{equation}

or

\begin{equation}
e_{3}=-\frac{1}{12}.
\end{equation}

Using eqs.(23), (10) and (97) that give $e_{3}$ in terms of $s$ and $k^{2}$,
and after a little algebra, we find the simple equation that relates $s$ to $%
k^{2}$ on $e=1$ to be given by eq.(51).

This simple equation (51) between $s$ and $k^{2}$ for $e=1$ can be used for
any $0\leq k^{2}\leq 1$. For example, we find that for $k^{2}=1$, $s=1/4=0.25
$, and for $k^{2}=1/2$, $s=1/(3\sqrt{2})=0.235702$. For $e=1$, and for small 
$k^{2}$, we have $s^{2}\simeq k^{2}/4$ which is a special case of $%
k^{2}\simeq 4es^{2}$ that is valid more generally for $0\leq e\leq 1$.

In addition, we find $g_{2}=1/12-s^{2}=1/12-k^{2}/[4(1+k^{2})^{2}]$ and
therefore

\begin{equation}
g_{2}=\frac{1-k^{2}+k^{4}}{12(1+k^{2})^{2}},
\end{equation}

and from the expression for $e_{1}$ and $e_{2}$ given by eq.(98), we find

\begin{eqnarray}
e_{1} &=&\frac{2-k^{2}}{12(1+k^{2})},  \notag \\
e_{2} &=&\frac{-1+2k^{2}}{12(1+k^{2})}.
\end{eqnarray}

Thus from the expressions (27), (30), and (21) for $q_{\min }$, $q_{1}$, and 
$\gamma $, we find that when $e=1$, they have the simple expressions given
by eqs.(53), (54) and (52).

Also, substituting eq.(48) into eq.(100) shows that $\varepsilon =1$ when $%
e=1$, i.e. $\varepsilon $ and $e$ coincide at $e=1$.

\section{Appendix C Some Simple Relations for the Special Case of $k^{2}=1/2$%
}

It is known that in elliptic functions, the squared modulus $k^{2}=1/2$ is a
special value for which many simple relations arise. We first consider the
case of $k^{2}=1/2$ in Region I. We note that substituting $k^{2}=1/2$ into
eq.(22) gives a relation $e_{1}+e_{3}=2e_{2}$, and substituting the
expressions of $e_{1},e_{2},e_{3}$ from eq.(23) into this relation gives $%
\theta /3=\pi /6$, or $\theta =\pi /2$. Thus $\cos \theta =0$ which results
in

\begin{equation}
g_{3}=0
\end{equation}

from eq.(24), which in turn gives a simple relationship

\begin{equation}
s^{2}=\frac{1}{6(1-e^{2})}\left( 1-\sqrt{\frac{1+2e^{2}}{3}}\right) .
\end{equation}

For example, we have $s=\sqrt{(3-\sqrt{3})/18}=0.265408$ for $e=0$, and $%
s=1/(3\sqrt{2})=0.235702$ for $e=1$ (using L'Hospital rule). These two
values represent the two terminal coordinates of the constant $k^{2}=1/2$
line in Region I (see Fig.1). The other solution of eq.(106) is eq.(58)
which is applicable for Region II as we shall show later in this Appendix.

We also find from eq.(23) that

\begin{eqnarray}
e_{1} &=&-e_{3}=\frac{\sqrt{g_{2}}}{2},  \notag \\
e_{2} &=&0,
\end{eqnarray}

and

\begin{equation}
\gamma =\sqrt[4]{g_{2}},
\end{equation}

and the orbit equation (20) becomes

\begin{equation*}
\frac{1}{q}=\frac{1}{3}-2\sqrt{g_{2}}cn^{2}(\gamma \phi ,1/\sqrt{2}),
\end{equation*}

where the $s$ value for $g_{2}$ in this case is given by eq.(107) for Region
I. The precession angle $\Delta \phi $ can be found from eq.(25) and from $%
K(1/\sqrt{2})=1.85407$. It is given by

\begin{equation*}
\frac{\Delta \phi }{\pi }=\frac{1.18034}{\sqrt[4]{g_{2}}}-2.
\end{equation*}

From eqs.(26)-(28) and (108), we find

\begin{equation}
\varepsilon =\frac{6(e_{2}-e_{3})}{1+6(e_{2}+e_{3})}=\frac{3\sqrt{g_{2}}}{1-3%
\sqrt{g_{2}}},
\end{equation}

which can be inverted and solved for $s$ in terms of $\varepsilon $, giving

\begin{equation}
s^{2}=\frac{3+6\varepsilon -\varepsilon ^{2}}{36(1+\varepsilon )^{2}}.
\end{equation}

Substituting eq.(111) into eq.(106) gives $e$ in terms of $\varepsilon $ as

\begin{equation}
e^{2}=1-\frac{12(1+\varepsilon )^{2}(1-\varepsilon )(1+3\varepsilon )}{%
(3+6\varepsilon -\varepsilon ^{2})^{2}}.
\end{equation}

Since for $e=0$, $s=\sqrt{(3-\sqrt{3})/18}$ when $k^{2}=1/2$ as we showed
above, substituting this $s$ value into eq.(106) gives $\varepsilon =(2/%
\sqrt{-3+2\sqrt{3}}-1)^{-1}=0.516588$. For $e=1$, $s=1/(3\sqrt{2})$ when $%
k^{2}=1/2$, and substituting this $s$ value into eq.(106) gives $\varepsilon
=1$ as it should. Thus eqs.(111) and (112) are the parametric equations for
the line of constant $k^{2}=1/2$ which can be used instead of eq.(107) as $%
\varepsilon $ takes the values between $0.516588$ and $1$.

Also, substituting $e_{2}=0$ from eq.(108) into eq.(27) gives

\begin{equation}
q_{\min }=3
\end{equation}

independent of $e$ for $k^{2}=1/2$, as shown in Table 3.

We now consider $k^{2}=1/2$ in Region II. From eqs.(35) and (33), we have

\begin{equation}
a=A+B=0.
\end{equation}

From eq.(32), and from $A=-B$, and $A^{3}=-B^{3}$, we arrive again at
eq.(106) with the same $g_{3}$ given by eq.(10). This explains why we stated
after eq.(107) that the other solution of eq.(106) given by eq.(58) gives
the relation between $s$ and $e$ for Region II. The two equations (108) and
(58) that give simple relations between $s$ and $e$ in two different regions
and that arise as two different solutions of the same equation (106) show a
rather remarkable symmetry exhibited by the special case $k^{2}=1/2$.

It also follows from eqs.(32) and (34) that

\begin{equation}
A=\frac{1}{2}\left( \frac{-g_{2}}{3}\right) ^{1/2}=\frac{1}{12}\sqrt{%
12s^{2}-1},
\end{equation}

\begin{equation}
\gamma =\left( 3A^{2}\right) ^{1/4}=\frac{1}{2}\left[ \frac{1}{3}(12s^{2}-1)%
\right] ^{1/4},
\end{equation}

and that the orbit equation (36) becomes

\begin{equation}
\frac{1}{q}=\frac{1}{3}+4\gamma ^{2}\frac{1-cn(2\gamma \phi ,1/\sqrt{2})}{%
1+cn(2\gamma \phi ,1/\sqrt{2})},
\end{equation}

where the $s$ value for the above equations is given by eq.(58) for $0\leq
e\leq 1$.

Since $a=0$, the initial distance $q_{2}$ of the planet from the blackhole
is $q_{2}=3$ from eq.(37), independent of $e$, as shown in Table 10.

\section{Appendix D The Boundaries of Region II}

On the upper boundary $s_{2}^{2}=1/(1-e^{2})$ of Region II, the planet
starts from the Schwarzschild horizon given by $q=1$, which implies that for 
$\phi =0$, $1/q=1/3+4a$ from eq.(36), or

\begin{equation}
1=\frac{1}{3}+4a.
\end{equation}

Hence

\begin{equation}
a=A+B=\frac{1}{6}.
\end{equation}

Since from eqs.(32) and (17),

\begin{equation}
AB=\frac{1}{4}\left( g_{3}^{2}-\frac{\Delta }{27}\right) ^{1/3}=\frac{g_{2}}{%
12},
\end{equation}

we can conclude from eqs.(35) and (34) that $k^{2}$ and $\gamma $ are given
on the boundary $s_{2}^{2}=1/(1-e^{2})$ of Region II by eqs.(60) and (61).

In particular, for $e=0$, $s=1$, we find $\gamma =(5/16)^{1/4}=0.747674$, $%
k^{2}=(1-1/\sqrt{5})/2=0.276393$, and this is the minimum value of $k^{2}$
in Region II.

We now consider the right boundary $e=1$ of Region II. Consider the
unbounded orbit of a planet coming from infinity that requires, from
eq.(36), that

\begin{equation}
\frac{1}{3}+4a=0,
\end{equation}

or

\begin{equation}
a=-\frac{1}{12}.
\end{equation}

\bigskip

From

\begin{equation}
A+B=a=-\frac{1}{12},
\end{equation}

and

\begin{equation}
AB=\frac{g_{2}}{12},
\end{equation}

and from eqs.(34) and (35), we find eqs.(62) and (64) that give $k^{2}$ and $%
\gamma $ in terms of $s$ on the boundary $e=1$ of Region II.

\bigskip

Finally, there is a special case of $g_{2}=0$ or $s^{2}=1/12$ ($s=1/2\sqrt{3}%
=0.288675135$) in Region II that is somewhat interesting. If $g_{2}$ of
eq.(10) is equal to zero, then $\Delta =27g_{3}^{2}$ from eq.(17), and from
eqs.(32)-(35), we find $A=(2g_{3})^{1/3}/2$, $B=0$, $a=A$, $\gamma
=(3A^{2})^{1/4}$, and noting that $A<0$ for $s^{2}=1/12$ and $0\leq e\leq 1$%
, we have

\begin{equation*}
k^{2}=\frac{1}{2}-\frac{\sqrt{3}A}{4\mid A\mid }=\frac{1}{2}+\frac{\sqrt{3}}{%
4}=0.933012702
\end{equation*}

which is independent of $e$, i.e. the constant $k^{2}=(2+\sqrt{3})/4$ curve
in Region II just above the boundary curve of Regions I and II in Fig.2 is a
horizontal line. Thus the terminating orbits represented by eq.(36) for any
point along this horizontal line are represented by elliptic functions of
the same squared modulus given above. It should be remembered, however, that
the orbits and the initial positions of the planet (the initial trajectory
of which is perpendicular to the line joining it to the blackhole) depend
also on $\gamma $ and $a$ that are dependent on the value of $e$. Thus the
orbit for $e=1$, for example, that gives $a=-1/12$ (and $\gamma =1/(2\cdot
3^{1/4})=0.379917843$) is a terminating orbit for a planet that is initially
infinitely far away from a blackhole.

\includepdf[pages={37-},
            ]{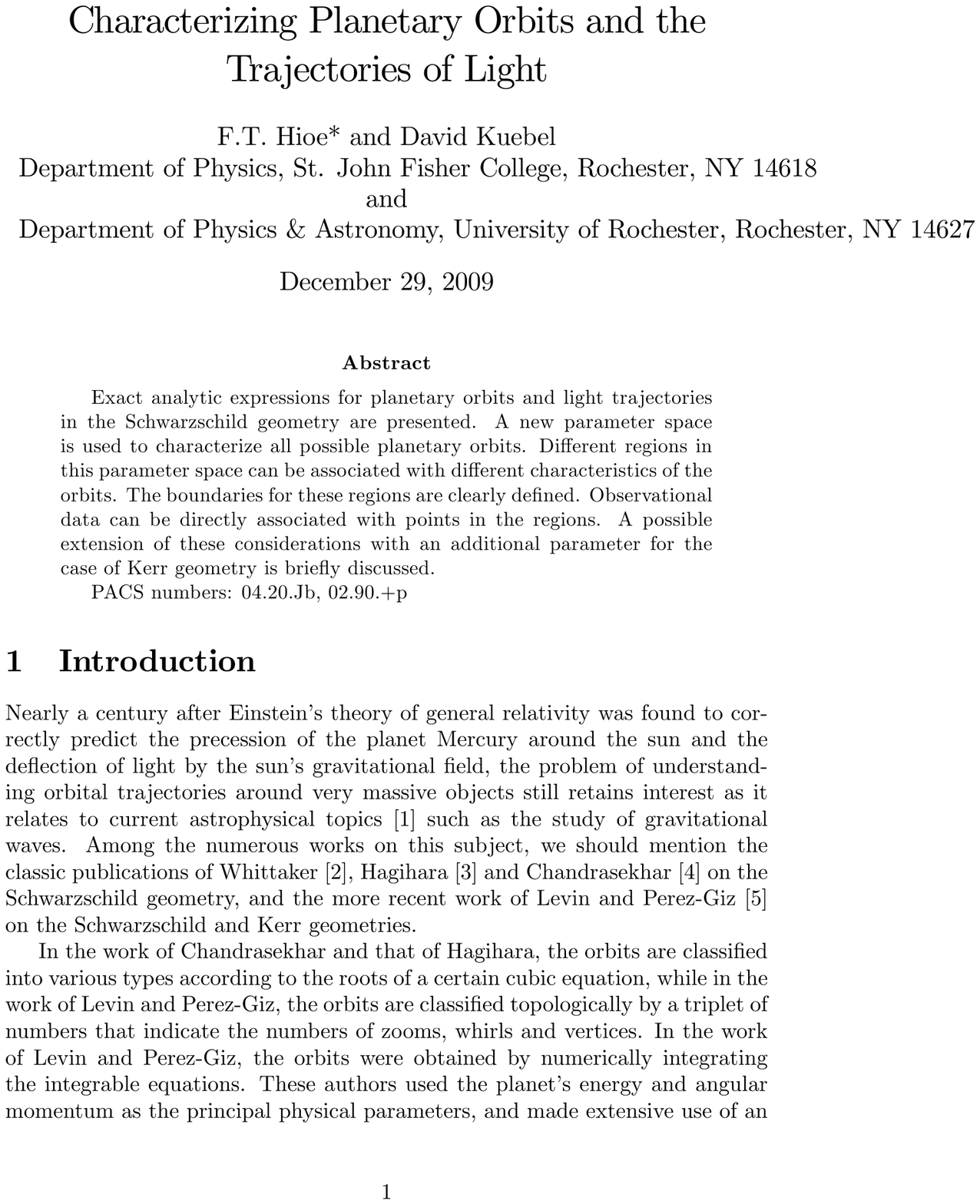}

\end{document}